\documentclass[pra,twocolumn,showpacs,nofootinbib,floatfix,superscriptaddress]{revtex4}

\usepackage{amsmath,amsfonts,amssymb,bm}
\usepackage{dcolumn}
\usepackage{braket}
\usepackage[final]{graphicx}
\usepackage{bm,tabularx}

\renewcommand{\Re}{\mathfrak{Re}}

\begin{document}

\title{Optimal quantum control of Bose-Einstein condensates in
  magnetic microtraps: Comparison of GRAPE and Krotov optimization
  schemes} 

\author{Georg J\"ager}
\affiliation{Institut f\"ur Physik, Karl-Franzens-Universit\"at Graz,
  Universit\"atsplatz 5, 8010 Graz, Austria} 

\author{Daniel M. Reich}
\author{Michael H. Goerz}
\author{Christiane P. Koch}
\affiliation{Theoretische Physik, Universit\"at Kassel,
  Heinrich-Plett-Str. 40, 34132 Kassel, Germany}

\author{Ulrich Hohenester}
\affiliation{Institut f\"ur Physik, Karl-Franzens-Universit\"at Graz,
  Universit\"atsplatz 5, 8010 Graz, Austria} 

\date{September 11, 2014}

\begin{abstract}
We study optimal quantum control of the dynamics of trapped Bose-Einstein
condensates: The targets are to split a condensate, residing initially in
a single well, into a double well, without inducing excitation; and to excite
a condensate from the ground to the first excited state of a single well. The
condensate is described in the mean-field approximation of the Gross-Pitaevskii
equation. We compare two optimization approaches in terms of their performance
and ease of use, namely gradient ascent pulse engineering (GRAPE) and Krotov's method. Both approaches are derived from the variational principle but
differ in the way the control is updated, additional costs are accounted for,
and second order derivative information can be included. 
We find that GRAPE produces smoother control fields and works in a black-box manner, whereas Krotov with a suitably chosen step size parameter converges faster but can produce sharp features in the control fields.
\end{abstract}


\maketitle

\section{Introduction}

Controlling complex quantum dynamics is a recurring theme in many
different areas of AMO physics and physical chemistry. Recent examples
include quantum state preparation~\cite{SayrinNat11,buecker:11},  
interferometry~\cite{vanfrank:14} and
imaging~\cite{LapertSciRep12,HaeberlePRL13}, 
or reaction control~\cite{RybakPRL11,GonzalezPRA12}. The central idea
of quantum control is to employ external fields to steer the dynamics
in a desired way~\cite{RiceBook,ShapiroBook}. The fields that
realize the desired dynamics can be determined by optimal
control theory (OCT) \cite{Tannor92,WerschnikJPB07}. 
An expectation value that encodes the target is then taken to be a
functional of the external field which is minimized or maximized. 
The target can be simply a desired final state~\cite{Tannor92}, or a
unitary operator~\cite{JosePRL02}, a
prescribed value of energy or position~\cite{DoriaPRL11}, or an
experimental signal such as a pump-probe trace~\cite{KaiserJCP04}. 

The  algorithms that can be employed for optimizing the target
functional broadly
fall into two categories -- those where changes in the field are
determined solely by evaluating the functional, such as simplex
algorithms~\cite{DoriaPRL11,CanevaPRA11}, 
and those that utilize derivative information, such as Krotov's
method~\cite{Konnov99,sklarz:02} or 
gradient ascent pulse engineering (GRAPE)~\cite{KhanejaJMR05}, possibly
combined with quasi-Newton methods~\cite{MachnesPRA11,eitan:12}.  
The solutions that one obtains typically do not only depend on 
the target functional but also on the specific algorithm that is
employed and the initial guess field. This is due to the fact that 
numerical optimization is always a local search which may find one of
possibly many optimal solutions or get stuck in a local
extremum. It is thus important to understand which features of an
optimal control solution are due to the optimization procedure and
which reflect truly physical properties of the quantum system. 

For example, when seeking to identify, by use of optimal control
theory, the  quantum
speed limit, i.e., the shortest possible time in which a quantum
operation can be carried out~\cite{CanevaPRL09}, the answer should be
independent of the algorithm. Moreover, in view of employing 
calculated solutions in an experiment, conditions such as limited
power, limited time resolution, or limited bandwidth need to be
met. The way in which the various optimization approaches can
accommodate such requirements differ greatly. 

Here, we study control of a Bose-Einstein condensate in a magnetic
microtrap, comparing several variants of a GRAPE-type
algorithm~\cite{hohenester.pra:07,hohenester.cpc:14a}
with Krotov's method~\cite{Konnov99,sklarz:02,reich:12}. We consider
two targets -- 
splitting the condensate, which resides initially in the ground state of
a single well, into a double well, without inducing excitation, and 
exciting the condensate from the ground to the first excited state of
a single well. The latter is important for stimulated processes in
matter waves, whereas the former presents a crucial step in
interferometry~\cite{shin:04,schumm:05,grond.njp:10}. A challenging
aspect of controlling a condensate is the non-linearity of the equation of
motion which can compromise or even prevent convergence of the
optimization~\cite{sklarz:02}. The two methods tackle this problem in
different ways, GRAPE by computing the search direction for new control fields
within the framework of Lagrange parameters and submitting the optimal control
search to generic minimization routines~\cite{peirce:88,hohenester.pra:07},
Krotov's method by accounting for the non-linearity of the equations of 
motion in the monotonicity conditions when constructing the algorithm~\cite{Konnov99,sklarz:02,reich:12}. 
Furthermore, the methods differ in the way in which additional
requirements such as smoothness of the control can be accounted
for. We compare the two optimization approaches with respect to the
solutions they yield as well as  their
performance, and ease of use.  Our study extends an earlier comparison of GRAPE-type algorithms with Krotov's method~\cite{MachnesPRA11} that was concerned with the linear 
Schr\"odinger equation and with finite-size (spin-type) quantum systems. 

Our paper is organized as follows:  
After introducing the equation of motion for the condensate dynamics
together with the control targets in
Sec.~\ref{sec:theory}, we briefly review the two optimization schemes in
Sec.~\ref{sec:octmethods}.  Section~\ref{sec:results}
presents our results for wavefunction splitting and shaking. Moreover,
we investigate the influence of the nonlinearity, 
the performance of the two algorithms, and the smoothness of the optimized
control  in Secs.~\ref{subsec:nonlin}
to~\ref{subsec:smooth}.   Our conclusions are presented in
Sec.~\ref{sec:conclusions}.

\section{Model and Optimization Problem}
\label{sec:theory}

In this paper we consider a quasi-1D condensate residing in a magnetic
confinement potential $V(x,\lambda(t))$ that can be controlled by some external
\textit{control parameter} $\lambda(t)$
\cite{hohenester.pra:07,buecker:11,buecker:13,hohenester.cpc:14a}.  We describe
the condensate dynamics within the mean-field framework of the Gross-Pitaevskii
equation, where $\psi(x,t)$ is the condensate wavefunction, normalized to one, whose time evolution
is governed by \cite{leggett:01} ($\hbar=1$) 
\begin{equation}\label{eq:gp}
 i\frac{\partial\psi(x,t)}{\partial t}=
   \Bigl(-\frac{1}{2M}\frac{\partial^2}{\partial x^2} + V(x,\lambda(t)) + 
   \kappa\bigl|\psi(x,t)\bigr|^2\Bigr)\psi(x,t)\,.
\end{equation}
The first term on the right-hand side is the operator for the kinetic
energy, the second one is the confinement potential, and the last term
is the nonlinear atom-atom interaction in the mean field
approximation.  $M$ is the atom mass and $\kappa$ is the strength of
the nonlinear atom-atom interactions, which is related to the effective one-dimensional interaction strength $U_0$ and the number of atoms $N$ through $\kappa=U_0(N-1)$ \cite{grond.pra:09b}.

We can now formulate our optimal control problem.  Suppose that the
condensate is initially described by the wavefunction
$\psi(x,0)=\psi_0(x)$ and the potential is varied in the time interval
$[0,T]$.  We are now seeking for an optimal time variation of
$\lambda(t)$ that brings the terminal wavefunction $\psi(x,T)$ as
close as possible to a \textit{desired} wavefunction $\psi_d(x)$.  To
rate the success for a given control, we introduce the cost function 
\begin{equation}\label{eq:statecost}
  J_T(\psi(T))=\frac 12\left[1-\left| \left< \psi_d |\psi(T)\right> \right |^2 \right]\,,
\end{equation}
which becomes zero when the terminal wavefunction matches the desired
one up to an arbitrary phase.  Optimal control theory aims at a
$\lambda_{\rm OCT}(t)$ that minimizes Eq.~\eqref{eq:statecost}.

\section{Optimization Methods}
\label{sec:octmethods}

\begin{table*}
  \centering
  \begin{ruledtabular}
  \begin{tabular}{lccccccl}
  Algorithm               & line search  &  free parameter          & deriv & penalty
  & penalty equation & update  &   update equation   \\ 
  \colrule
  GRAPE grad L2 & yes &  $\gamma$    & 1    & $\dot\lambda^2$ &  Eq.~\eqref{eq:cost} & concurrent & Eq.~\eqref{eq:L2} \\ 
  GRAPE grad H1 & yes &  $\gamma$    & 1    & $\dot\lambda^2$ &  Eq.~\eqref{eq:cost} & concurrent & Eq.~\eqref{eq:H1} \\   
  GRAPE BFGS L2 & yes &  $\gamma$    & 2    & $\dot\lambda^2$ &  Eq.~\eqref{eq:cost} & concurrent & Eq.~\eqref{eq:L2} \\ 
  GRAPE BFGS H1 & yes &  $\gamma$    & 2    & $\dot\lambda^2$ &  Eq.~\eqref{eq:cost} & concurrent & Eq.~\eqref{eq:H1} \\ 
  Krotov                  & no  &  $k$         & 1    &
  $(\Delta\lambda)^2$  & Eq.~\eqref{eq:controlkrotov}        & sequential & Eq.~\eqref{eq:controlkrotov2} \\ 
  KBFGS                   & no  &  $k$         & 2    &
  $(\Delta\lambda)^2$  &  Eq.~\eqref{eq:controlkrotov}  & sequential & Eq.~(38) of Ref.~\cite{eitan:12} \\ 
  \end{tabular}
  \end{ruledtabular}
  \caption{Optimization approaches used in this paper. For each algorithm, we
  specify whether a line search is used; what free parameter is
  available to influence the optimization; the order of the derivative for the determination of the new control parameter; the
  penalty term that is added to Eq.~\eqref{eq:statecost}, with 
  $\Delta \lambda = \lambda - \lambda_{\rm ref}$; the equation for the cost function; the type of the update of the control; and the update equation used in our simulations. 
  } \label{tab:algos}
\end{table*}

In this paper, we apply two different optimal control approaches, namely
a gradient ascent pulse engineering (GRAPE) scheme~\cite{KhanejaJMR05} and
Krotov's method~\cite{Konnov99,sklarz:02,reich:12}, which will be discussed
separately below. An overview of the control approaches is given in
Table~\ref{tab:algos}.

\subsection{GRAPE: Functional and Optimization Scheme}
\label{subsec:grape}

The GRAPE scheme for Bose-Einstein condensates has been presented in detail
elsewhere~\cite{hohenester.pra:07,buecker:13,jaeger.pra:13,hohenester.cpc:14a},
for this reason we only briefly introduce the working equations.
Experimentally, strong variations of the control parameter are difficult to
achieve.  Therefore, we add to the cost function an additional term
\cite{hohenester.pra:07,borzi:08,vonwinckel:08}, 
\begin{equation}\label{eq:cost}
  J(\psi(T),\lambda)=J_T(\psi(T))+\frac{\gamma}{2} \int_0^T[\dot{\lambda}(t)]^2 dt\,.
\end{equation}
Mathematically, the additional term penalizes strong variations of
the control parameter and is needed to make the OCT problem 
well-posed~\cite{hohenester.pra:07,borzi:08,vonwinckel:08}.
Through $\gamma$ it is possible to weight the relative importance of
wavefunction matching and control smoothness.  Below we will set $\gamma\ll 1$
such that $J$ is dominated by the terminal cost $J_T$.

In order to bring the system from the initial state $\psi_0$ to the
terminal state $\psi(T)$ we have to fulfill the Gross-Pitaevskii
equation, which enters as a \textit{constraint} in our optimization
problem.  The constrained optimization problem can be turned into an
unconstrained one by means of Lagrange multipliers $p(x,t)$, whose
time evolution is governed by~\cite{hohenester.pra:07} 
\begin{equation}\label{eq:adjoint}
  i\dot p=\left(-\frac{1}{2M}\frac{\partial^2}{\partial x^2} + 
  V(x,\lambda(t))+2\kappa|\psi|^2\right)p+\kappa\psi^2 p^*\,,
\end{equation}
subject to the terminal condition
$p(T)=i\langle\psi_d|\psi(T)\rangle\psi_d$.  The optimal control
problem is then composed of the Gross-Pitaevskii equation~\eqref{eq:gp} and
Eq.~\eqref{eq:adjoint}, which must be fulfilled simultaneously
together with \cite{hohenester.pra:07}
\begin{equation}\label{eq:optcontrol}
  \gamma\ddot\lambda=
  -\Re\bigl<p\bigr|\frac{\partial V}{\partial\lambda}\bigl|\psi\bigr>
\end{equation}
for the optimal control.  This expression differs from standard
GRAPE~\cite{KhanejaJMR05} and results from minimizing changes in the
control, cf. Eq.~\eqref{eq:cost}. 

This set of equations can be also employed for a non-optimal control where
Eq.~\eqref{eq:optcontrol} is not fulfilled.  In this case Eq.~\eqref{eq:gp} is
solved forwards in time and
Eq.~\eqref{eq:adjoint} backwards in time, and the search direction
$\nabla_\lambda J$ for an improved control is calculated from one of the
equations \cite{hohenester.pra:07,vonwinckel:08,hohenester.cpc:14a} 
\begin{eqnarray}
  \nabla_\lambda J\phantom{]}\!\!&=&\!\!
  -\gamma\ddot\lambda-
  \Re\,\bigl<p\bigr|\frac{\partial V}{\partial\lambda}\bigl|\psi\bigr> \,\,\,
  \mbox{for $L^2$ norm}\label{eq:L2}\qquad\\
  -\frac{d^2 }{dt^2}\bigl[\nabla_\lambda J\bigr]\!\!&=&\!\!
  -\gamma\ddot\lambda-
  \Re\,\bigl<p\bigr|\frac{\partial V}{\partial\lambda}\bigl|\psi\bigr> \,\,\,
  \mbox{for $H^1$ norm.}\label{eq:H1}\qquad
\end{eqnarray}
These two expressions are obtained by interpreting, on the right hand
side of Eq.~\eqref{eq:cost}, the integral 
$\int_0^T[\dot\lambda]^2\,dt=\langle \dot\lambda,\dot\lambda\rangle_{L^2}=
\langle \lambda,\lambda\rangle_{H^1}$
in terms of an $L^2$ or $H^1$
norm~\cite{vonwinckel:08,grond.pra:09b}. The $H^1$ norm implies 
that one additionally has to solve a Poisson equation, see
the derivative operator on the left hand side of
Eq.~\eqref{eq:H1}. This generally results in a much smoother 
time dependence of the control parameters while the additional
numerical effort for solving the Poisson equation is negligible.  As
for an optimal control, both Eqs.~\eqref{eq:L2} and 
\eqref{eq:H1} yield $\nabla_\lambda J=0$.   

Here we solve the optimal control equations using the Matlab toolbox
\textsc{octbec} \cite{hohenester.cpc:14a}.  The ground and desired states of the Gross-Pitaevskii equation are computed using the optimal damping algorithm \cite{dion:07,hohenester.cpc:14a}.
The control parameters are obtained
iteratively using either a conjugate gradient method (GRAPE grad), which
only uses first order information, or a quasi-Newton BFGS
scheme~\cite{bertsekas:99} (GRAPE BFGS), which also takes into account
second-order information via an approximated Hessian. In both cases, the
optimization employs a line search to determine the optimal step size in the
direction of a given gradient. The pulse update is calculated for all time
points simultaneously, making the GRAPE schemes \emph{concurrent}.

\subsection{Krotov's method: Functional and Optimization scheme}
\label{subsec:krotov}

Krotov's method~\cite{Konnov99} provides an alternative optimal control
implementation.  The main idea is to add to Eq.~\eqref{eq:statecost} a
vanishing term~\cite{Konnov99,sklarz:02,reich:12}, which is chosen
such that the minimum of the new function is also a minimum of $J$.
However, for non-optimal $\lambda(t)$ one can devise a scheme that
always gives a new control corresponding to a lower cost
function.  Thus, Krotov's method leads to a monotonically convergent
optimization algorithm that is expected to exhibit much faster
convergence. 

Our implementation closely follows
Refs.~\cite{sklarz:02,reich:12,eitan:12}.  Specifically, the cost reads
\begin{equation}\label{eq:costkrotov}
  J(\psi(T),\lambda)
  = J_T(\psi(T))+
    \int_0^T \frac{[\lambda(t)-\lambda_{\rm ref}(t)]^2}{S(t)}  dt\,,
\end{equation}
where the reference field $\lambda_{\rm ref}(t)$
is typically chosen to be the control from the previous
iteration~\cite{PalaoPRA03}. The second term in
Eq.~\eqref{eq:costkrotov} penalizes changes in the control from one
iteration to the next, and ensures that as an optimum is approached the value of
the functional is increasingly determined by only $J_T$. 
$S(t)=k s(t)$ is a shape function that
controls the turning on and off of the control fields, $k$ is a step size
parameter, and $s(t)\in[0,1]$ is bound between $0$ and $1$. 

Let $\psi^{(i)}(t)$ and $\lambda^{(i)}(t)$ denote the wavefunction and control
parameter, respectively, in the $i$th iteration of the optimal control loop.
To get started, we first solve for an initial guess $\lambda^{(0)}(t)$ the
Gross-Pitaevskii equation \eqref{eq:gp} and the adjoint equation \eqref{eq:adjoint} 
for the co-state $p(t)$, which is backward-propagated in time with the same terminal condition as in GRAPE,
in order to obtain $\psi^{(0)}(t)$ and $p^{(0)}(t)$.  In the next step, we
solve the Gross-Pitaevskii equation \textit{simultaneously} with the equation
for the new control field 
\begin{widetext}
\begin{equation}
\label{eq:controlkrotov}
\lambda^{(i+1)}(t) = \lambda^{(i)}(t)
  + S(t)  \Re\, \Braket{p^{(i)}(t)|
                        \left[
                        \frac{\partial V}{\partial \lambda}
                        \right]_{\lambda^{(i+1)}(t)}
                        | \psi^{(i+1)}(t)}
  +         \Re\, \frac{\sigma(t)}{2i}
            \Braket{\Delta \psi(t)|
                    \left[
                    \frac{\partial V}{\partial \lambda}
                    \right]_{\lambda^{(i+1)}(t)}
                    | \psi^{(i+1)}(t)}\,,
\end{equation}
\end{widetext}
where $\psi^{(i+1)}(t)$ is obtained by propagating $\psi(t=0)$ forward
in time using the updated pulse.~\footnote{The co-states of this work and of Ref.~\cite{reich:12} are related through $p=i\chi$.  With this definition the adjoint equation \eqref{eq:adjoint} and the terminal condition $p(T)$ are the same for GRAPE and Krotov.  As consequence, the scalar products on the right hand side of Eq.~\eqref{eq:controlkrotov} involve the real rather than the imaginary part.} 
The fact that $\psi^{(i+1)}(t)$ appears on the right hand side of the update
equation implies that the update at a given time $t$ depends on the updates at
all earlier times, making Krotov's method \emph{sequential}. This type of
update makes it non-straightforward to include a cost term on the
derivative of the control as in Eq.~\eqref{eq:cost}, since the derivative at
a given time $t$ requires knowledge of past \emph{and} future values of
$\psi(t)$.


The last term in Eq.~\eqref{eq:controlkrotov} 
with $\Delta\psi(t)=\psi^{(i+1)}(t)-\psi^{(i)}(t)$ is generally
needed to ensure convergence in presence of the nonlinear mean-field
term $\kappa|\psi(t)|^2$ of the Gross-Pitaevskii equation. Convergence is
achieved through a proper choice of  $\sigma(t)$ 
\cite{sklarz:02,reich:12}.  In this work we neglect this additional
contribution for simplicity, as it is of only minor importance for
the moderate $\kappa$ values of our present concern. 

\begin{figure}
\centerline{\includegraphics[width=\columnwidth]{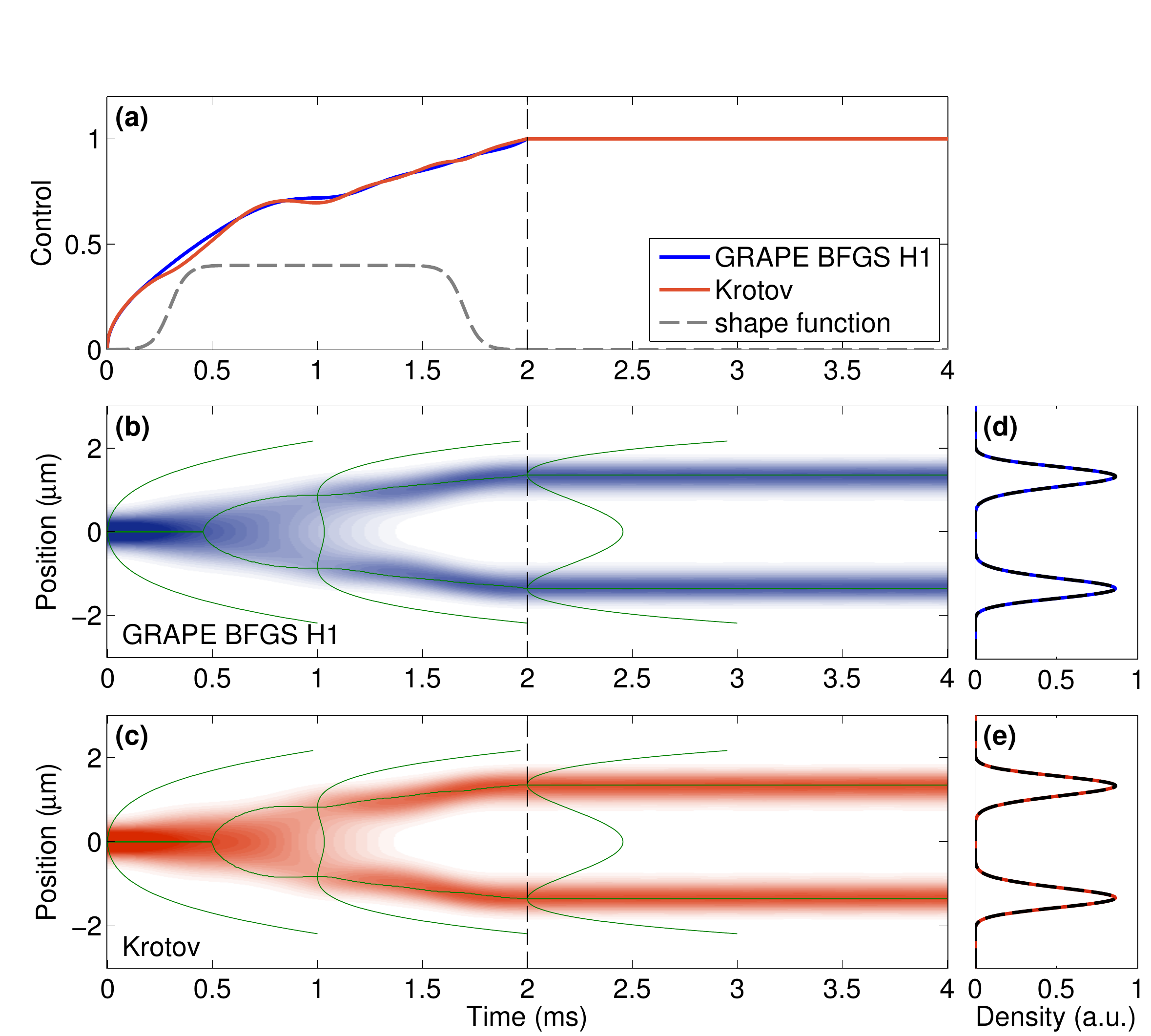}}
\caption{(Color online) Wavefunction splitting through the transformation of the
confinement potential from a single to a double well.  (a) The solid lines report the control parameters $\lambda(t)$ for the GRAPE and Krotov
optimizations, respectively.  The potential is held constant after the terminal
time $T=2$ ms.  The dashed line shows the shape function $s(t)$ of
Eq.~\eqref{eq:costkrotov} used in our version of Krotov's method,
scaled by a factor of $0.4$ for better visibility. (b,c) Density plots of the
condensate density $n(x,t)=|\psi(x,t)|^2$ during the splitting.  The solid
lines show the confinement potentials at three selected times and the time
variation of the potential minima.  (d,e) Terminal (solid lines) and desired
(dashed lines) densities, which are indistinguishable.  In the optimization we
set $\gamma=10^{-6}$ and $k=10^{-3}$.}
\label{fig:splitting1}
\end{figure}

\begin{figure}
\centerline{\includegraphics[width=\columnwidth]{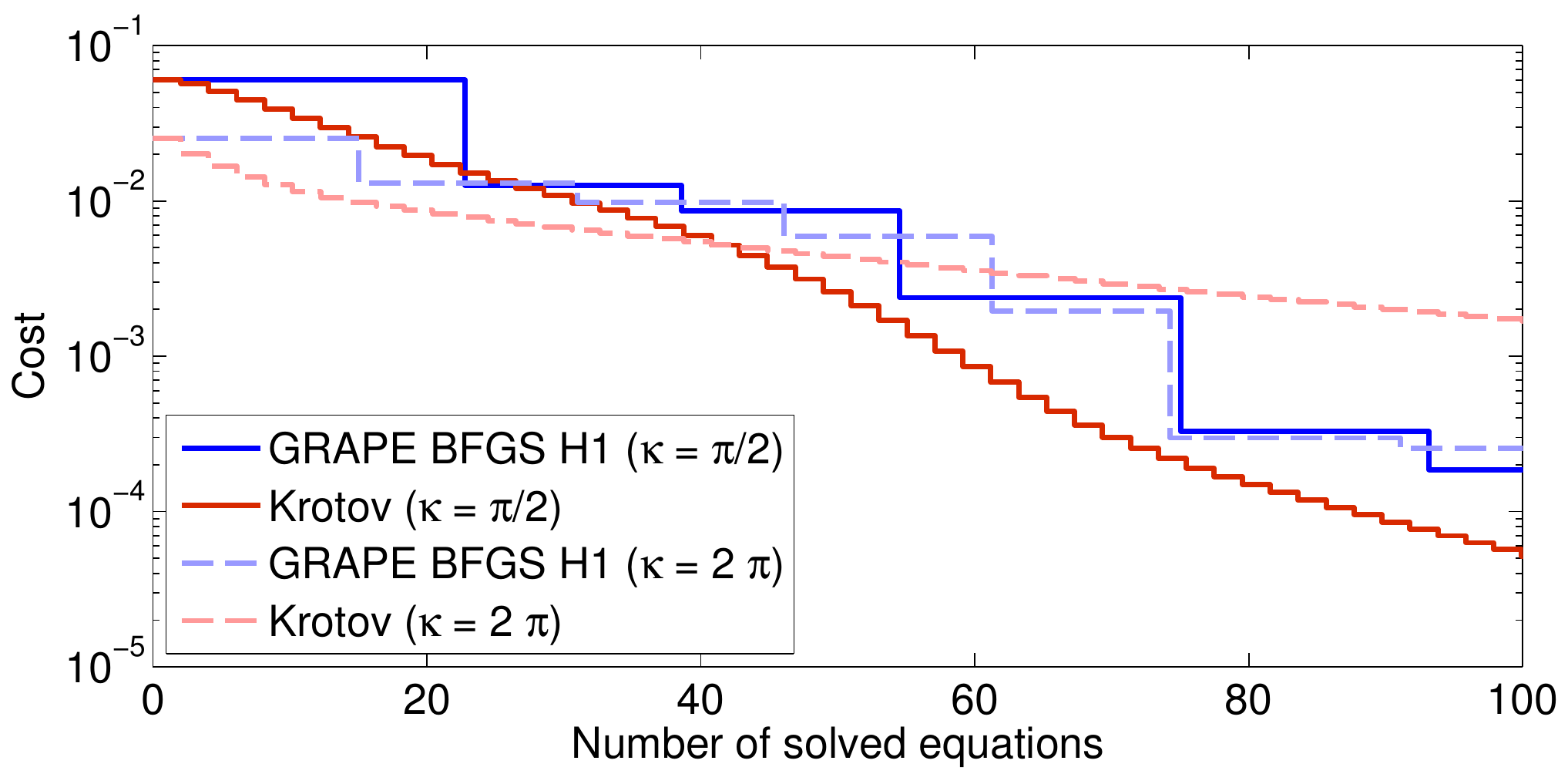}}
\caption{(Color online) Cost function versus number of solved equations (either Gross-Pitaevskii or adjoint equation) for
GRAPE and Krotov.  For GRAPE one optimization iteration consists of numerous
solutions of the Gross-Pitaevskii equation \eqref{eq:gp} during a line search,
which are followed by a solution of the adjoint equation \eqref{eq:adjoint} once
a minimum is found, to obtain a new search direction $\nabla_\lambda J$.  For
Krotov one optimization iteration consists of a Gross-Pitaevskii solution,
subject to Eq.~\eqref{eq:controlkrotov2}, and a subsequent solution of the
adjoint equation \eqref{eq:adjoint}.  In our simulations we use $k=10^{-3}$.
The dashed lines report results of simulations with a larger nonlinearity
$\kappa/\hbar=2\pi\times 1000$ Hz.  In the legend we report the $\kappa$ values
in units used in our simulations, with $\hbar=1$ and time measured in
milliseconds \cite{hohenester.cpc:14a}.}
\label{fig:splitting2}
\end{figure}

The derivative ${\partial V}/{\partial\lambda}$ in
Eq.~\eqref{eq:controlkrotov} has to be computed for $\lambda^{(i+1)}(t)$,
thus leading to an implicit equation for $\lambda^{(i+1)}(t)$.  When $k$
is chosen sufficiently small, such that the control parameter varies
only moderately from one iteration to the next, one can obtain the new
control fields approximately from 
\begin{equation}
\begin{split}
\label{eq:controlkrotov2}
\lambda^{(i+1)}(t) & \approx  \lambda^{(i)}(t) \\
    & + S(t) \, \Re \Braket{%
      p^{(i)}(t) | 
      \left[
      \frac{\partial V}{\partial \lambda}
      \right]_{\lambda^{(i)}(t)}
      | \Psi^{(i+1)}(t)
    }\,.
\end{split}
\end{equation}
Otherwise one can employ an iterative Newton scheme for the
calculation of $\lambda^{(i+1)}(t)$, as briefly described in
Appendix~\ref{sec:newtonkrotov}.  In all our simulations we found
Eq.~\eqref{eq:controlkrotov2} to provide sufficiently accurate
results. 
Once the new wavefunctions $\psi^{(i+1)}(t)$ and control parameters
$\lambda^{(i+1)}(t)$ are computed, we get the adjoint variables $p^{(i+1)}(t)$ through
the solution of Eq.~\eqref{eq:adjoint} and continue with the Krotov optimization
loop until the cost function $J$ is small enough or a certain number of
iterations is exceeded.

As a variant,  we also use a combination of Krotov's method with the
BFGS method (KBFGS) \cite{eitan:12}.
It includes an approximated Hessian via the Krotov gradient as an additional
term in the update equation~\eqref{eq:controlkrotov}. However, for technical
reasons and differently from the GRAPE BFGS algorithm, no line search is
employed.

\section{Results}
\label{sec:results}

In this paper, we consider two control problems. The first one is \textit{condensate
splitting}, where the condensate initially resides in one well which is
subsequently split into a double well.  In our simulations we employ the
confinement potential of Lesanovsky et al.~\cite{lesanovsky:06} where the
control parameter $\lambda(t)$ is associated with a radio frequency magnetic field \cite{hohenester.pra:07}.
The objective is to bring at the terminal time $T$ the condensate wavefunction
to the ground state of the double well potential. 

In the second control problem the condensate wavefunction is
excited from the ground to the first excited state of a single well potential.
The confinement potential is an anharmonic single-well
potential, details and a parameterized form of $V(x)$ can be found
in~\cite{buecker:11,buecker:13,hohenester.cpc:14a}.  The shakeup is achieved by
displacing the potential origin according to $V(x-\lambda(t))$, where
$\lambda(t)$ now corresponds to the position of the potential minimum, i.e.,
through \textit{wavefunction shaking}.  Experimental
realizations of such shaking protocols have been reported in
\cite{buecker:11,buecker:13,vanfrank:14}.
 
In our simulations GRAPE and Krotov start with the same initial guess. 
The terminal time is set to $T=2$ ms throughout. 
Unless stated differently, we use a nonlinearity
$\kappa/\hbar=2\pi\times 250$ Hz ($\kappa=\pi/2$ for
units with $\hbar=1$ and time measured in milliseconds, as used in our
simulations \cite{hohenester.cpc:14a}).

\subsection{Splitting vs.\ Shaking}
\label{subsec:splitting}

\begin{figure}[tb]
\centerline{\includegraphics[width=\columnwidth]{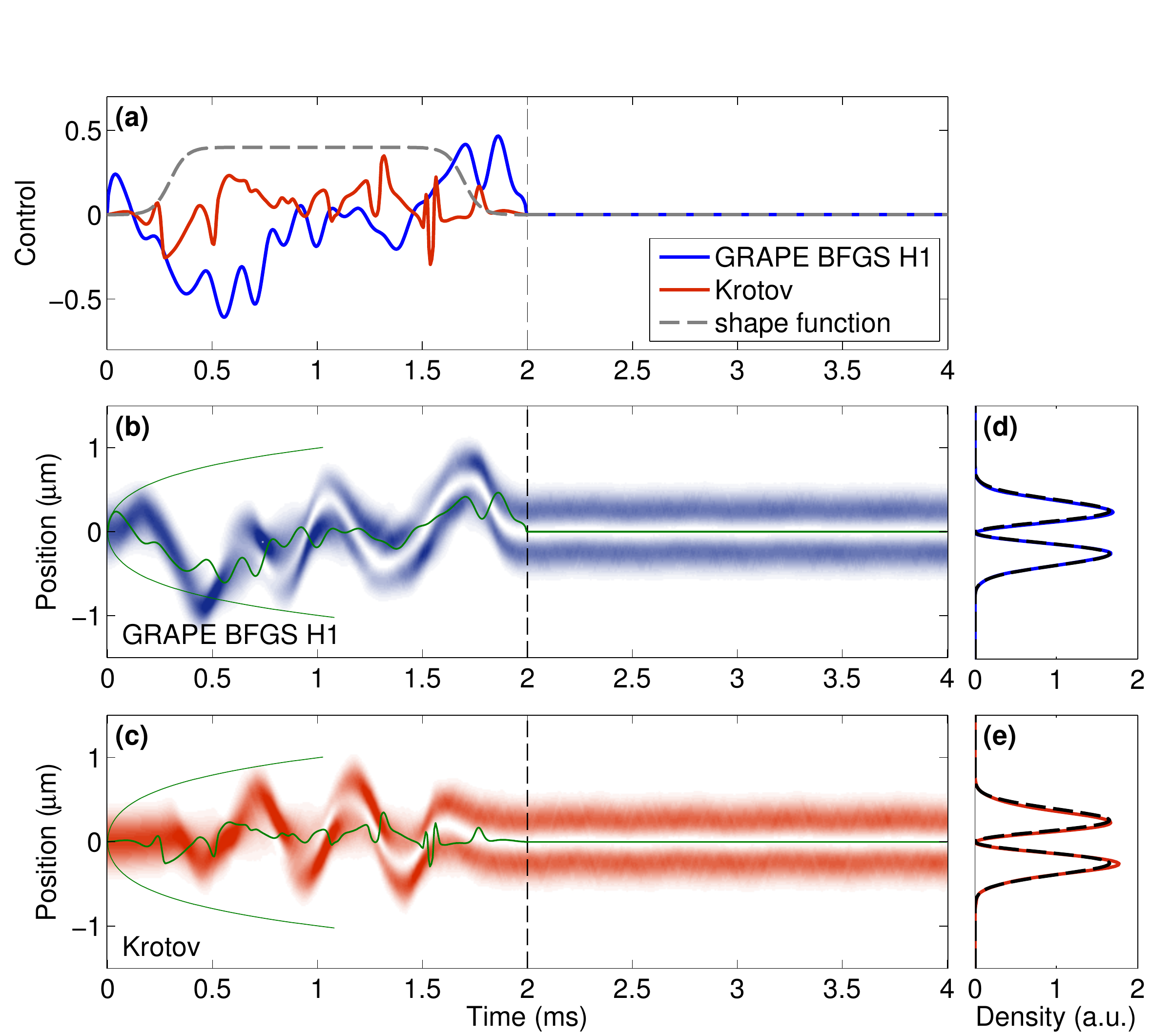}}
\caption{(Color online) Same as Fig.~\ref{fig:splitting1} but for shaking process.}
\label{fig:shaking1}
\end{figure}

Figure~\ref{fig:splitting1} shows (a) the controls obtained from our
GRAPE and Krotov optimizations for condensate splitting, together with (b,c) the density maps of the condensate wavefunction.  
The potential is held constant after the terminal
time $T=2$ ms of the control process.  Figs.~\ref{fig:splitting1}(d,e) show the
square moduli of the terminal (solid lines) and desired (dashed lines)
condensate wavefunctions, which are almost indistinguishable, thus demonstrating
the success of both control protocols.  This can be also seen from the density
maps which show no time variations at later times, when the potential is held
constant, in accordance to the fact that the terminal wavefunction is the
groundstate of the double well trap.

Figure~\ref{fig:splitting2} compares the efficiency of the GRAPE and Krotov
optimizations.  We plot the cost function $J_T$ versus the number $n$ of
equations solved during optimization.  For both GRAPE and Krotov, $n$ counts the
solutions of either the Gross-Pitaevskii or the adjoint equation.
The actual computer run
times depend on the details of the numerical implementation, but are comparable
for both schemes.  As can be seen in Fig.~\ref{fig:splitting2}, in the GRAPE
optimization the cost function decreases in large steps after a given number of
solved equations, whereas in the Krotov optimization $J_T$ decreases continuously.
The cost evolution of GRAPE can be attributed to the BFGS search algorithm,
where a line search is performed along a given search direction.  Once the
minimum is found, the step is accepted ($J_T$ drops) and a new search direction
is obtained through the solution of the adjoint equation.  In contrast, the
Krotov algorithm is constructed such that $J_T$ decreases monotonically in each
iteration step.  Altogether, GRAPE and Krotov optimizations
perform equally well.

\begin{figure}[tb]
\centerline{\includegraphics[width=\columnwidth]{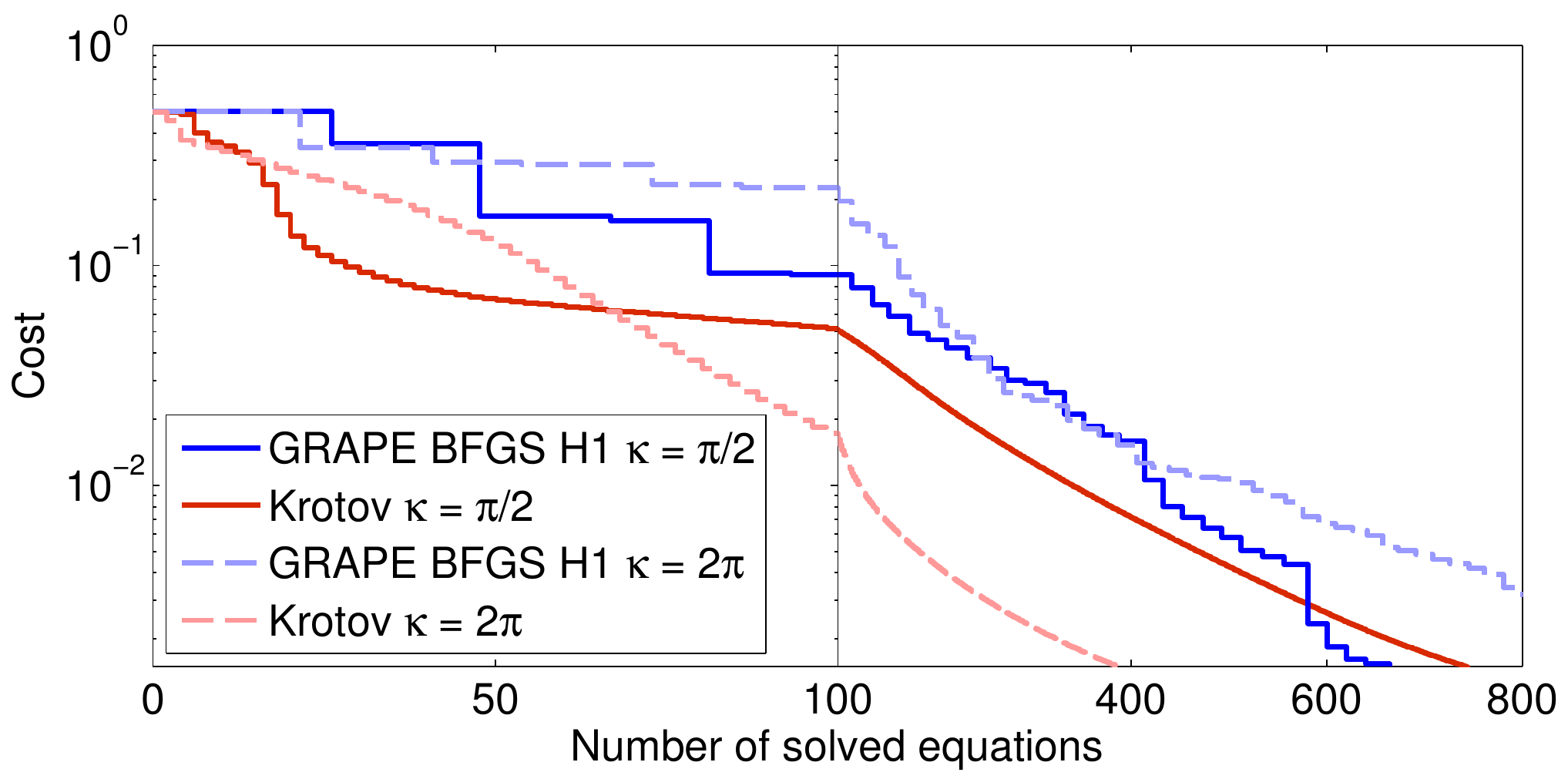}}
\caption{(Color online) Same as Fig.~\ref{fig:splitting2} but for shaking
process.  We use $k=5\times 10^{-3}$.}
\label{fig:shaking2}
\end{figure}

In comparison to condensate splitting, the shakeup process is a
considerably more complicated control problem.  
Figure~\ref{fig:shaking1} shows the optimized control parameters as well as the
time evolution of the condensate densities.  Both GRAPE and Krotov succeed
comparably well. Regarding the control fields, the GRAPE one is smoother
than the Krotov one, due to the penalty term on $\dot\lambda(t)$ in
Eq.~\eqref{eq:cost}.  From Fig.~\ref{fig:shaking2} we observe that a much higher
number of optimization iterations is needed, in comparison to wavefunction
splitting, for both optimization methods to significantly reduce $J_T$.
Initially, $J_T$ decreases more rapidly for the Krotov optimization, but after
a larger number $n$ of solved equations, say around $n\sim 600$, GRAPE
performs better.  

\subsection{Influence of Nonlinearity}
\label{subsec:nonlin}

We investigate the influence of the nonlinear atom-atom interaction on
the convergence of the optimization loop.  The dashed lines in
Fig.~\ref{fig:splitting2} report results for splitting simulations with a larger
nonlinearity $\kappa/\hbar=2\pi\times 1000$ Hz.  While the GRAPE convergence
depends only weakly on $\kappa$, Krotov converges significantly slower for
larger $\kappa$ values.  

Things are different for the shaking shown in Fig.~\ref{fig:shaking2}.
While the GRAPE performance again depends only weakly on $\kappa$, Krotov
converges \textit{faster} with increasing $\kappa$.
Because of the lack of a line search in the Krotov algorithm, the convergence
behavior is far more dependent on specific features of the control landscape
which depend strongly on $\kappa$.

\subsection{Convergence Behavior}
\label{subsec:convergence}

Next, we inquire into the details of the convergence properties for the
optimization of the shakeup process.  By comparing GRAPE with Krotov, we will
identify the advantages and disadvantages of the respective optimization
methods.  

\begin{figure}[tb]
\centerline{\includegraphics[width=0.9\columnwidth]{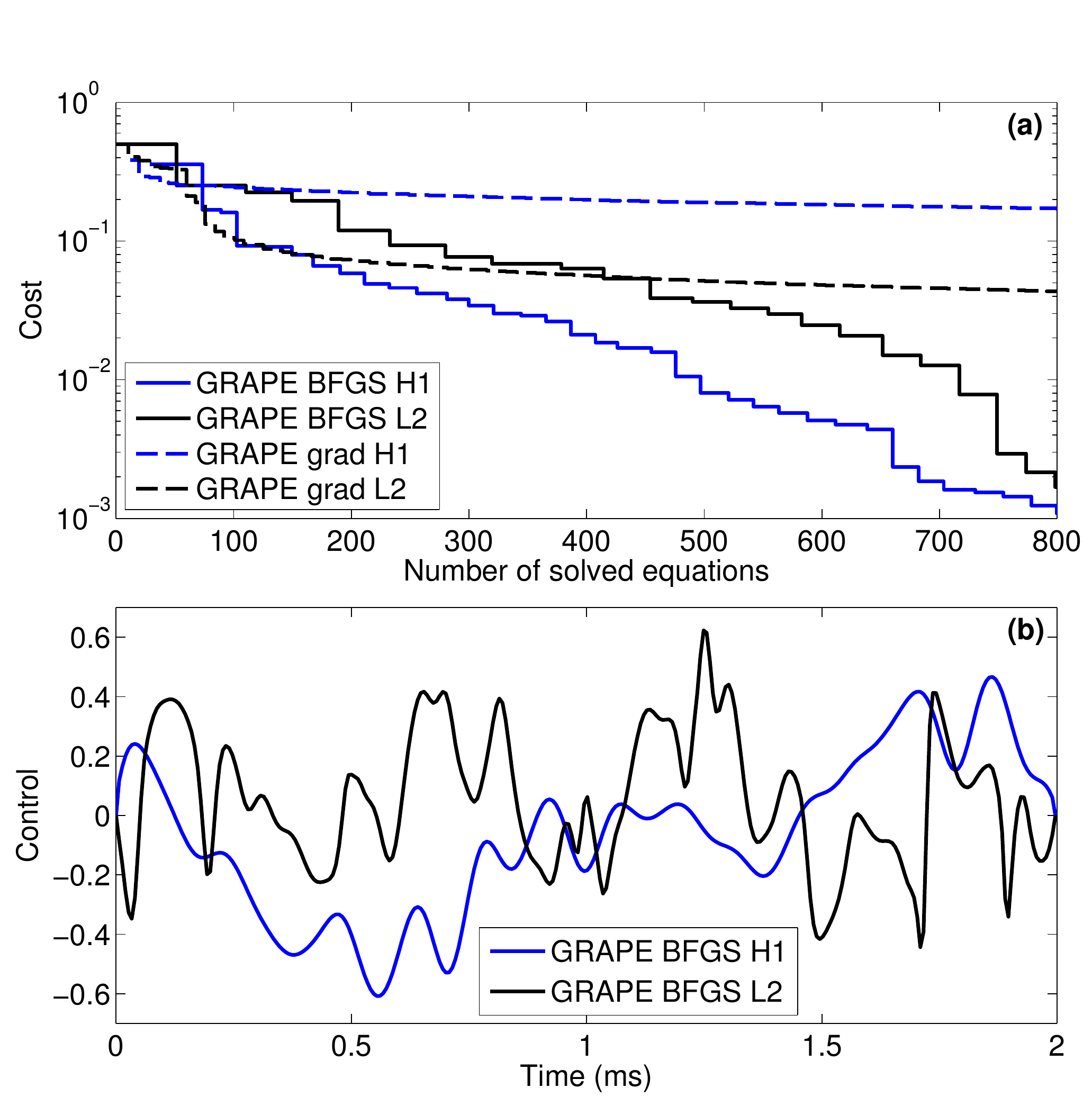}}
\caption{(Color online) (a) Cost function versus number of solved equations for
conjugate gradient (grad) and BFGS optimization schemes, and for search
directions obtained from Eqs.~(\ref{eq:L2},\ref{eq:H1}) with $H^1$ or $L^2$
norm, respectively.  (b) Optimal control parameters $\lambda(t)$ for BFGS
solutions.}
\label{fig:shakingGRAPE}
\end{figure}

Figure~\ref{fig:shakingGRAPE}(a) shows
the terminal cost function $J_T$ versus the number of  solved equations of motion $n$ for
the different GRAPE schemes.  It is evident that the conjugate gradient
solutions reach a plateau after a certain number of iterations.
In contrast, the BFGS solutions decrease significantly even at later stages of the optimization.  We attribute this behavior to the use of the second order derivative
information. The GRAPE BFGS scheme, which estimates the Hessian of $J$ in
addition to $\nabla_\lambda J$, can take larger steps to cross flat regions of
$J$, contrary to the (first-order) GRAPE gradient scheme, which gets stuck. 

Figure~\ref{fig:shakingGRAPE}(b) shows the control fields for the
GRAPE BFGS schemes.  Although both optimization strategies perform equally well, the solutions
obtained with $H^1$ norm are smoother and probably better suited for
experimental implementation.   

\begin{figure}[tb]
\centerline{\includegraphics[width=0.9\columnwidth]{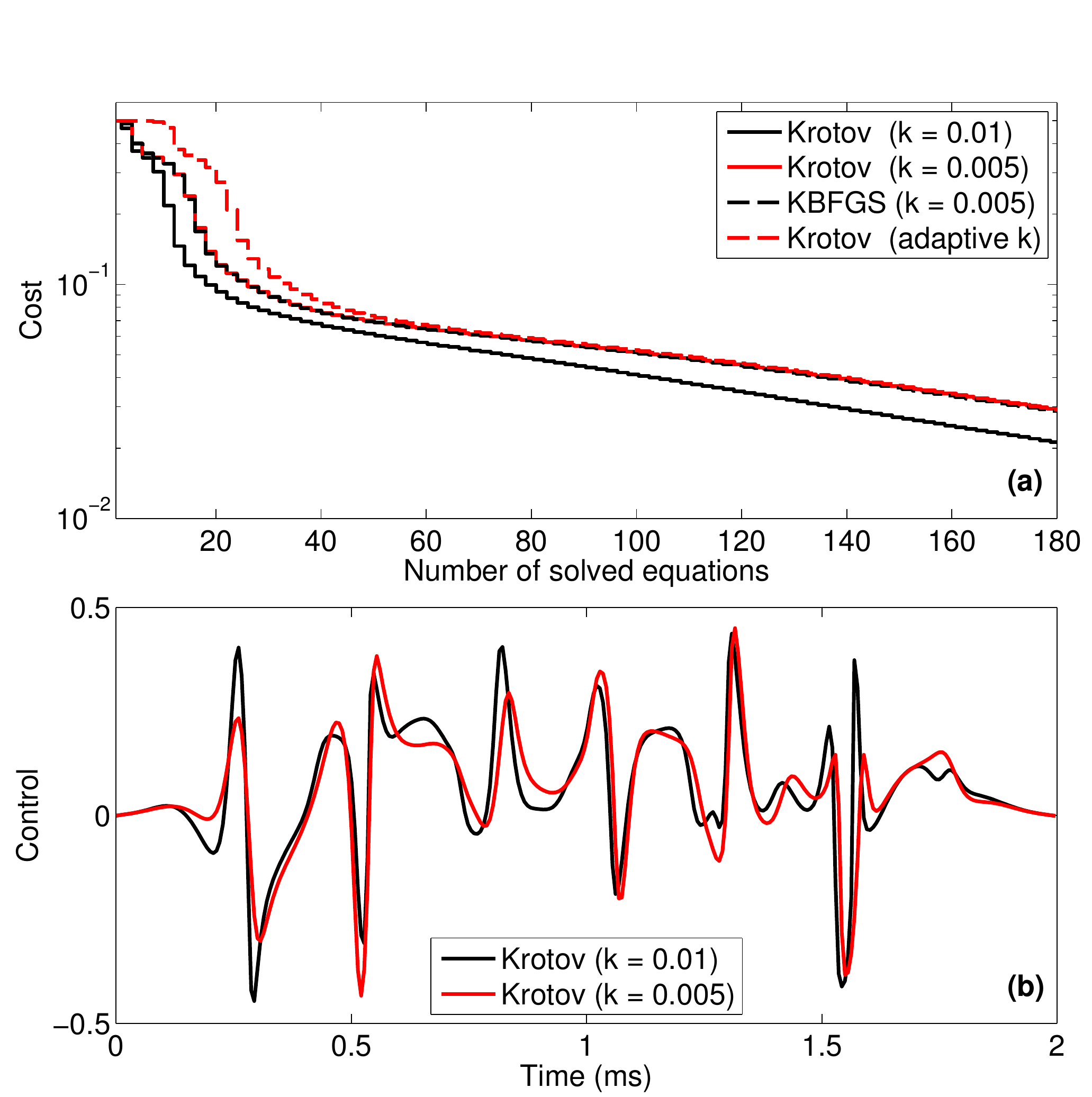}}
\caption{(Color online) Same as Fig.~\ref{fig:shakingGRAPE} but for Krotov
optimization.  We investigate the influence of different mixing parameters $k$
between the old and new control fields, see Eq.~\eqref{eq:controlkrotov2}, as
well as a scheme with an adaptive $k$ choice (see text for details).  KBFGS
reports the optimization result for a combination of Krotov's method
with BFGS~\cite{eitan:12}.} 
\label{fig:shakingKrotov}
\end{figure}

Figure~\ref{fig:shakingKrotov} presents (a) $J_T$ versus $n$ and (b) the control
parameters for the Krotov optimization.  The solid line with $k=0.005$ in
panel (a) is identical to the one shown in Fig.~\ref{fig:shaking2}.  When we
increase $k$ (black line) the cost function drops more rapidly.  However, we
found that larger $k$ values can lead to sharp variations in $\lambda(t)$ which
might be problematic for experimental implementations, as will be discussed in
more detail below.

\begin{figure*}
\centerline{\includegraphics[width=1.5\columnwidth]{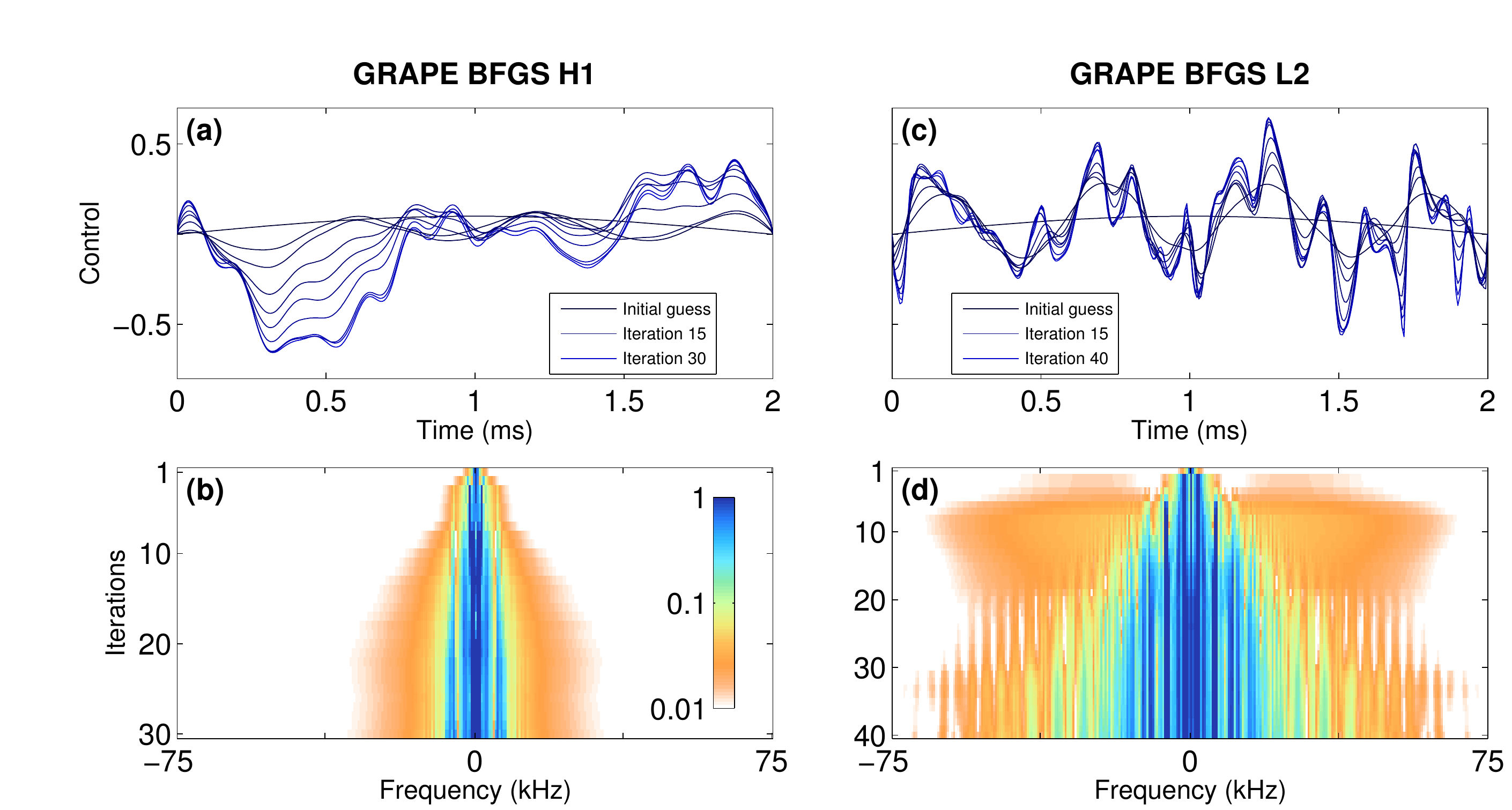}}
\caption{(Color online) (a,c) Evolution of control parameters during the
optimization process for GRAPE.  (b,d) Density plot of power spectra of the
control parameters displayed in panels (a,c).  We use a logarithmic color scale.  In panels (b) and (d) the numbers of iterations are chosen such that the final cost function $J_T$ becomes approximately $10^{-2}$, the numbers of solved equations (see Figs.~2 and 4) are approximately (b) 500 and (d) 700.}
\label{fig:performanceGRAPE}
\end{figure*}

\begin{figure*}
\centerline{\includegraphics[width=1.5\columnwidth]{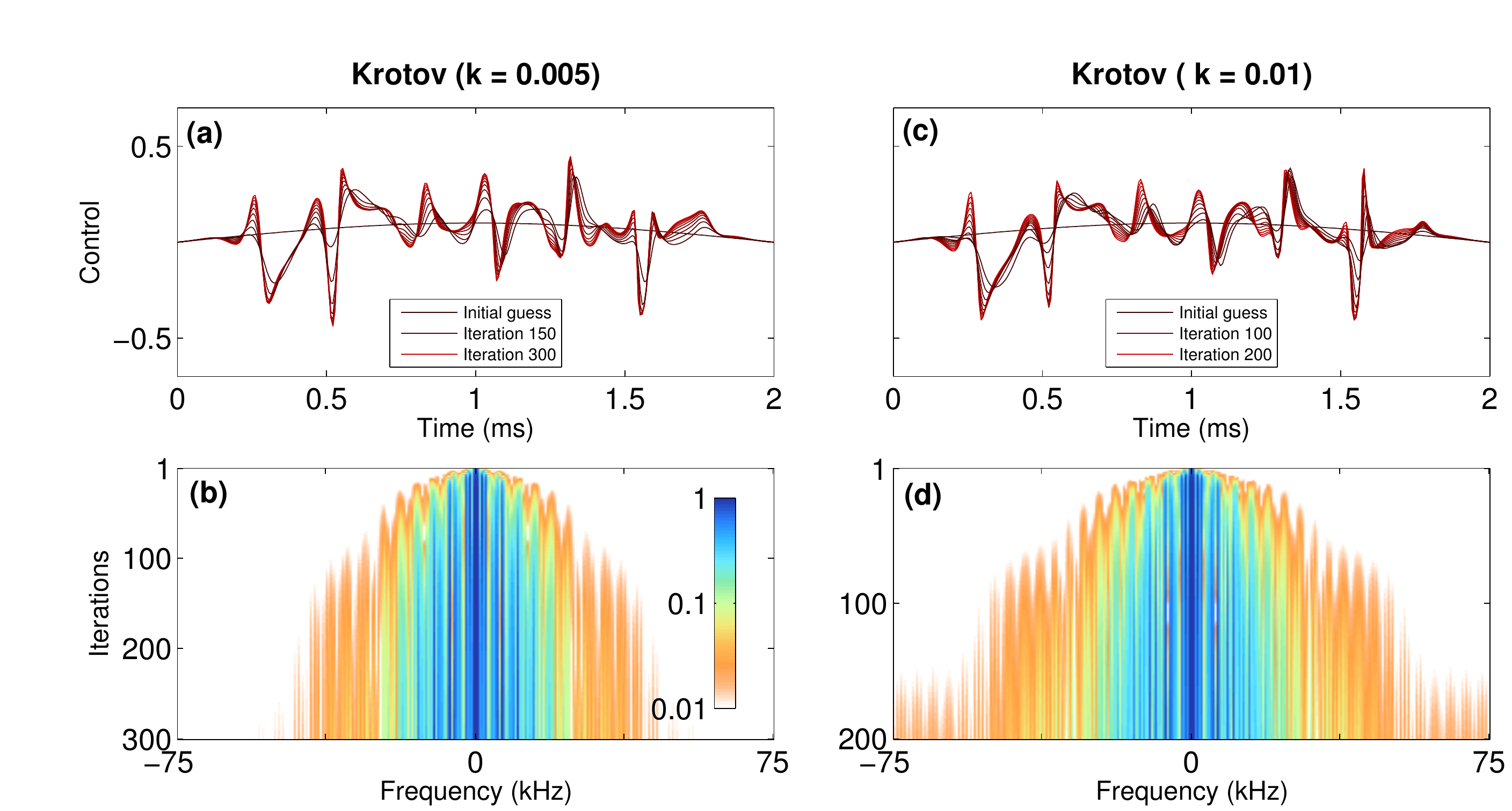}}
\caption{(Color online) Same as Fig.~\ref{fig:performanceGRAPE} but for Krotov
optimization.  The numbers of solved equations are (b) 600 and (d) 400.}
\label{fig:performanceKrotov}
\end{figure*}

In Fig.~\ref{fig:shakingKrotov} we additionally display results for a
simulation using a combination of Krotov's method with the BFGS scheme
(KBFGS)~\cite{eitan:12}.  The performance of KBFGS is similar to the
simpler optimization procedure of Eq.~\eqref{eq:controlkrotov2}, a
finding in accordance with 
Ref.~\cite{eitan:12}. 
We attribute this to the fact that within the
Krotov scheme only a small portion of the control landscape is
explored, because the monotonic convergence enforces small control
updates, in contrast to GRAPE where larger regions are scanned by the
line search.  As consequence, the improvement in the 
Krotov search direction via the Hessian is minimal.

Finally, the dashed line for adaptive $k$ shows results for an optimization that starts
with a small $k$  value, which subsequently increases in each iteration until the cost
decreases by a desired amount (here 2.5 per cent) within one iteration.  This $k$
value is then kept constant for the rest of the optimization.  The idea behind this
strategy is that the choice of $k$ is crucial for convergence, but the optimal
value is different for each problem.  Generally, finding a suitable value for
$k$ requires some trial and error.

\subsection{Features of the Control}
\label{subsec:smooth}

For many experimental implementations it is indispensable to use smooth control
parameters.  In the following we investigate the smoothness of the optimal
controls obtained by the different optimization methods. 

Figure~\ref{fig:performanceGRAPE}(a) shows for GRAPE BFGS H1 the evolution of the $\lambda(t)$
values during optimization.  One observes that during the first few
iterations the characteristic features of $\lambda(t)$ emerge, which then become
refined in the course of further iterations.  Fig,~\ref{fig:performanceGRAPE}(b)
reports the power spectra (square moduli of Fourier transforms) of the
$\lambda(t)$ history during optimization.  During the first say 20 iterations
the Fourier-transformed control parameter $\tilde\lambda(\nu)$ spectrally
broadens, indicating the emergence of sharp features during optimization.  With
increasing iterations the spectral width of $\tilde\lambda(\nu)$ remains
approximately constant.

Results of the GRAPE BFGS L2 optimization are shown in Figs.~\ref{fig:performanceGRAPE}(c,d).  We
observe that, in contrast to the H1 results, $\lambda(t)$ acquires sharp
features during optimization, as also reflected by the broad power spectrum.
This is because initially the gradient $\nabla_\lambda J$, which determines the
search direction for improved control parameters, exhibits strong variations.
These variations are washed out in the H1 optimization through the solution
of the Poisson equation, see Eq.~\eqref{eq:H1}, leading to significantly
smoother control parameters.

In GRAPE, the user must additionally provide the weighting factor $\gamma$ of
Eq.~\eqref{eq:cost} that determines the relative importance of terminal cost and
control smoothness.  For the problems under study, we found that the performance
of GRAPE does not depend sensitively on the value of $\gamma$, and we usually
use a small value such that the cost is dominated by the terminal cost.

Figs.~\ref{fig:performanceKrotov}(a,c) show the $\lambda(t)$ history during
a Krotov optimization for different step sizes $k$, and panels (b,d) report the
corresponding power spectra.  In comparison to the GRAPE BFGS H1 optimization, the power
spectra are significantly broader, in particular for the larger $k$ values.  This is
due to the fact that in the functional used for the Krotov optimization there
is no penalty term that enforces smoothness of the control (and thus a narrow
spectrum).

The choice of the step size $k$ is rather critical for the Krotov performance.  With
increasing $k$ the cost function decreases more rapidly during optimization.
However, values of $k$ that are too large can lead to numerical instabilities.
These instabilities result from the discretization of the update equation;
mathematically, Krotov is only guaranteed to converge monotonically for any
value of $k$ if the control problem is continuous. 



\begin{figure}[tb]
\centerline{\includegraphics[width=\columnwidth]{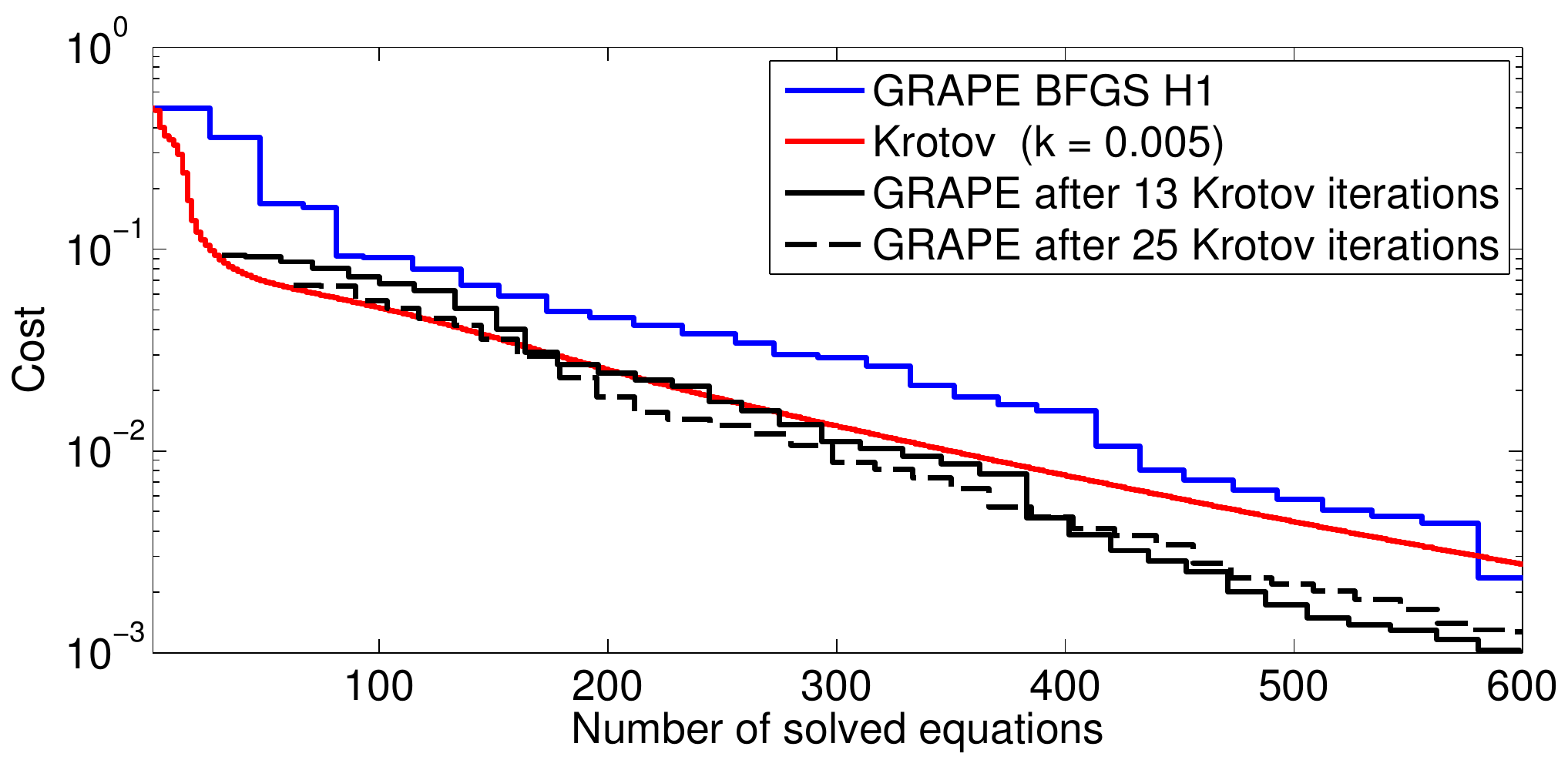}}
\caption{(Color online) Same as Fig.~\ref{fig:shaking2}, but for a combined
GRAPE--Krotov scheme where one initially starts with the Krotov method and
switches to GRAPE after a given number of iterations.}
\label{fig:GRAPEKrotov}
\end{figure}

One might wonder whether a combination of both approaches would give the best of
two worlds.  In Fig.~\ref{fig:GRAPEKrotov} we present results for simulations
where we start with a Krotov optimization and switch to GRAPE after a given
number of iterations.  As can be seen, the performance of this combined
optimization does not offer a particular advantage over
genuine GRAPE or Krotov optimizations.  This is probably due to
differences between  the optimal
control fields $\lambda(t)$ obtained by the two approaches, such that 
$\lambda(t)$ needs to be significantly modified when changing from one scheme to
the other.  In addition, the BFGS search algorithm of GRAPE uses the information
of previous iterations in order to estimate the Hessian of the control space,
and this information is missing when changing schemes.

\section{Conclusions and Outlook}\label{sec:conclusions}

Based on the two examples investigated in the previous section, namely
wavefunction splitting and shaking in a magnetic microtrap, we now set out to
analyze the advantages and disadvantages of the GRAPE and Krotov optimization
methods which are tied to the functional that is minimized in each case. 

First, when the optimization converges fast to an optimal solution, such as for wavefunction splitting
investigated in Sec.~\ref{subsec:splitting}, both optimization algorithms
perform equally well, even without carefully tuning the free parameters $\gamma$ or $k$.  For such problems, the choice of algorithm is a matter of
personal preference.  On the other hand, for optimization problems with slow convergence, such as wavefunction shaking, more care has to be taken.
Specifically, there are significant differences between the two algorithms in
terms of free parameters vs.\ speed of convergence as well as possible cost
functionals vs.\ features of the obtained optimal control.

While GRAPE BFGS utilizes a line search to ensure monotonic convergence and to
obtain the optimal step size in each iteration, the speed of convergence in
Krotov's method is mainly determined by the free parameter $k$. On the one hand this
means that GRAPE BFGS works better ``out of the box'' since it automatically
determines the best step size in each step. On the other hand,
the convergence is slowed down due to the necessity of a line search.

It is also evident from our results that both algorithms yield controls with
features that can be understood in terms of additional costs
introduced in the
functional. For GRAPE we use a cost that penalizes a large derivative of
the control which results in smooth controls in the end. For Krotov's
method we employ
a penalty on changes in the amplitude of the control in each iteration.
Correspondingly this leads to controls that have a smaller integrated
intensity and come at the cost of a less smooth control.

In principle it is conceivable to modify the Krotov algorithm to take into
account an additional cost term on the derivative of the control. While we
conjecture that this will lead to controls that are comparable with those
obtained in the GRAPE framework, the necessary modification of the Krotov
algorithm is beyond the scope of the current work.

In the context of controlling Bose-Einstein condensates with experimentally
smooth controls, the optimization with the GRAPE BFGS method, a functional
enforcing smoothness, and use of the $H^1$ norm appears to be the method of
choice.  It is a black--box scheme with practically no problem-dependent
parameters, it gives the desired smooth control fields, and works for various
nonlinearity parameters $\kappa$.

In contrast, the Krotov optimization without an appropriate penalty term in the
functional can converge faster but usually also leads to sharp features in
the control.  A sensitive choice of the step size $k$ is indispensable to
achieve a compromise between fast convergence and smoothness.  If smoothness is
not an issue or extremely fast convergence is needed, the Krotov method is
preferable.  

A combination of GRAPE and Krotov in the sense of switching from
one method to the other during the optimization did not result in any
significant gain. This is explained by the different control solutions
that are found by the different methods which do not easily facilitate
a transition between them. It points to the fact that many control
solutions exist and which solution is identified by the optimization depends
strongly on the additional constraints~\cite{JosePRA13} as well as the
optimization method.

\section*{Acknowledgments}

This work has been supported in part by the Austrian science fund FWF
under project P24248, by NAWI Graz, and the European Union under Grant 
No. 297861 (QUAINT).

\begin{appendix}

\section{}\label{sec:newtonkrotov}

In this appendix we briefly show how to numerically solve the equation
\begin{widetext}
\begin{equation}\label{eq:controlkrotov3}
  \lambda^{(i+1)}(t) =\lambda^{(i)}(t)
  +S(t)\,\Re\,
  \bigl<p^{(i)}(t)\bigr|\left[\frac{\partial V}{\partial\lambda}\Bigl|_{\lambda^{(i+1)}(t)}\right]
  \bigl|\psi^{(i+1)}(t)\bigr>\,,
\end{equation}
which differs from Eq.~\eqref{eq:controlkrotov2} in that the potential
derivative is evaluated for $\lambda^{(i+1)}(t)$.  Things can be easily generalized
for the additional $\sigma(t)$ term of Eq.~\eqref{eq:controlkrotov}.  Let
$\lambda_0(t)$ denote an initial guess for the solution of
Eq.~\eqref{eq:controlkrotov3}, e.g.\ the solution of
Eq.~\eqref{eq:controlkrotov2}.  We now set
$\lambda^{(i+1)}(t)=\lambda_0(t)+\delta\lambda(t)$, where $\delta\lambda(t)$ is assumed to
be a small quantity.  Thus, we can expand the second term on the right hand side
of Eq.~\eqref{eq:controlkrotov3} in lowest order of $\delta\lambda(t)$ to obtain
\begin{equation}\label{eq:controlkrotov4}
  \lambda_0(t)+\delta\lambda(t)\approx\lambda^{(i)}(t)+S(t)\,\Re\,
  \bigl<p^{(i)}(t)\bigr|\left[\frac{\partial V}{\partial\lambda}\Bigl|_{\lambda_0(t)}
  +\frac{\partial^2 V}{\partial\lambda^2}\Bigl|_{\lambda_0(t)}\delta\lambda(t)
  \right]\bigl|\psi^{(i+1)}(t)\bigr>\,.
\end{equation}
Separating the contributions of $\delta\lambda$ from the rest, we get
\begin{equation}
  \left(1-S(t) \,\Re\,
  \bigl<p^{(i)}(t)\bigr|\left[
  \frac{\partial^2 V}{\partial\lambda^2}\Bigl|_{\lambda_0(t)}
  \right]\bigl|\psi^{(i+1)}(t)\bigr>\right)\delta\lambda(t)\approx
  -(\lambda_0(t)-\lambda^{(i)}(t))+
  S(t)\bigl<p^{(i)}(t)\bigr|\left[\frac{\partial V}{\partial\lambda}\Bigl|_{\lambda_0(t)}\right]
  \bigl|\psi^{(i+1)}(t)\bigr>\,,
\end{equation}
\end{widetext}
%
which can be solved for $\delta\lambda(t)$.  If $|\delta\lambda(t)|<\varepsilon$ is
smaller than some small tolerance $\varepsilon$, we set
$\lambda^{(i+1)}(t)\to\lambda_0(t)+\delta\lambda(t)$.  Otherwise we set
$\lambda_0(t)\to\lambda_0(t)+\delta\lambda(t)$ and repeat the Newton iteration until
convergence.  Typically only few iterations are needed to reach tolerances of
the order of $\varepsilon=10^{-6}$.

\end{appendix}


\begin{thebibliography}{39}
\expandafter\ifx\csname natexlab\endcsname\relax\def\natexlab#1{#1}\fi
\expandafter\ifx\csname bibnamefont\endcsname\relax
  \def\bibnamefont#1{#1}\fi
\expandafter\ifx\csname bibfnamefont\endcsname\relax
  \def\bibfnamefont#1{#1}\fi
\expandafter\ifx\csname citenamefont\endcsname\relax
  \def\citenamefont#1{#1}\fi
\expandafter\ifx\csname url\endcsname\relax
  \def\url#1{\texttt{#1}}\fi
\expandafter\ifx\csname urlprefix\endcsname\relax\def\urlprefix{URL }\fi
\providecommand{\bibinfo}[2]{#2}
\providecommand{\eprint}[2][]{\url{#2}}

\bibitem[{\citenamefont{Sayrin et~al.}(2011)\citenamefont{Sayrin, Dotsenko,
  Zhou, Peaudecerf, Rybarczyk, Gleyzes, Rouchon, Mirrahimi, Amini, Brune
  et~al.}}]{SayrinNat11}
\bibinfo{author}{\bibfnamefont{C.}~\bibnamefont{Sayrin}},
  \bibinfo{author}{\bibfnamefont{I.}~\bibnamefont{Dotsenko}},
  \bibinfo{author}{\bibfnamefont{X.}~\bibnamefont{Zhou}},
  \bibinfo{author}{\bibfnamefont{B.}~\bibnamefont{Peaudecerf}},
  \bibinfo{author}{\bibfnamefont{T.}~\bibnamefont{Rybarczyk}},
  \bibinfo{author}{\bibfnamefont{S.}~\bibnamefont{Gleyzes}},
  \bibinfo{author}{\bibfnamefont{P.}~\bibnamefont{Rouchon}},
  \bibinfo{author}{\bibfnamefont{M.}~\bibnamefont{Mirrahimi}},
  \bibinfo{author}{\bibfnamefont{H.}~\bibnamefont{Amini}},
  \bibinfo{author}{\bibfnamefont{M.}~\bibnamefont{Brune}},
  \bibnamefont{et~al.}, \bibinfo{journal}{Nature}
  \textbf{\bibinfo{volume}{477}}, \bibinfo{pages}{73–77}
  (\bibinfo{year}{2011}).

\bibitem[{\citenamefont{B{\"u}cker et~al.}(2011)\citenamefont{B{\"u}cker,
  Grond, Manz, Berrada, Betz, Koller, Hohenester, Schumm, Perrin, and
  Schmiedmayer}}]{buecker:11}
\bibinfo{author}{\bibfnamefont{R.}~\bibnamefont{B{\"u}cker}},
  \bibinfo{author}{\bibfnamefont{J.}~\bibnamefont{Grond}},
  \bibinfo{author}{\bibfnamefont{S.}~\bibnamefont{Manz}},
  \bibinfo{author}{\bibfnamefont{T.}~\bibnamefont{Berrada}},
  \bibinfo{author}{\bibfnamefont{T.}~\bibnamefont{Betz}},
  \bibinfo{author}{\bibfnamefont{C.}~\bibnamefont{Koller}},
  \bibinfo{author}{\bibfnamefont{U.}~\bibnamefont{Hohenester}},
  \bibinfo{author}{\bibfnamefont{T.}~\bibnamefont{Schumm}},
  \bibinfo{author}{\bibfnamefont{A.}~\bibnamefont{Perrin}}, \bibnamefont{and}
  \bibinfo{author}{\bibfnamefont{J.}~\bibnamefont{Schmiedmayer}},
  \bibinfo{journal}{Nature Phys.} \textbf{\bibinfo{volume}{7}},
  \bibinfo{pages}{508} (\bibinfo{year}{2011}).

\bibitem[{\citenamefont{va{n F}rank et~al.}(2014)\citenamefont{va{n F}rank,
  Negretti, Berrada, B{\"u}cker, Montangero, Schaff, Schumm, Calarco, and
  Schmiedmayer}}]{vanfrank:14}
\bibinfo{author}{\bibfnamefont{S.}~\bibnamefont{va{n F}rank}},
  \bibinfo{author}{\bibfnamefont{A.}~\bibnamefont{Negretti}},
  \bibinfo{author}{\bibfnamefont{T.}~\bibnamefont{Berrada}},
  \bibinfo{author}{\bibfnamefont{R.}~\bibnamefont{B{\"u}cker}},
  \bibinfo{author}{\bibfnamefont{S.}~\bibnamefont{Montangero}},
  \bibinfo{author}{\bibfnamefont{J.~F.} \bibnamefont{Schaff}},
  \bibinfo{author}{\bibfnamefont{T.}~\bibnamefont{Schumm}},
  \bibinfo{author}{\bibfnamefont{T.}~\bibnamefont{Calarco}}, \bibnamefont{and}
  \bibinfo{author}{\bibfnamefont{J.}~\bibnamefont{Schmiedmayer}},
  \bibinfo{journal}{Nature Commun.} \textbf{\bibinfo{volume}{5}},
  \bibinfo{pages}{4009} (\bibinfo{year}{2014}).

\bibitem[{\citenamefont{Lapert et~al.}(2012)\citenamefont{Lapert, Zhang,
  Janich, Glaser, and Sugny}}]{LapertSciRep12}
\bibinfo{author}{\bibfnamefont{M.}~\bibnamefont{Lapert}},
  \bibinfo{author}{\bibfnamefont{Y.}~\bibnamefont{Zhang}},
  \bibinfo{author}{\bibfnamefont{M.~A.} \bibnamefont{Janich}},
  \bibinfo{author}{\bibfnamefont{S.~J.} \bibnamefont{Glaser}},
  \bibnamefont{and} \bibinfo{author}{\bibfnamefont{D.}~\bibnamefont{Sugny}},
  \bibinfo{journal}{Sci. Rep.} \textbf{\bibinfo{volume}{2}},
  \bibinfo{pages}{589} (\bibinfo{year}{2012}).

\bibitem[{\citenamefont{H\"aberle et~al.}(2013)\citenamefont{H\"aberle,
  Schmid-Lorch, Karrai, Reinhard, and Wrachtrup}}]{HaeberlePRL13}
\bibinfo{author}{\bibfnamefont{T.}~\bibnamefont{H\"aberle}},
  \bibinfo{author}{\bibfnamefont{D.}~\bibnamefont{Schmid-Lorch}},
  \bibinfo{author}{\bibfnamefont{K.}~\bibnamefont{Karrai}},
  \bibinfo{author}{\bibfnamefont{F.}~\bibnamefont{Reinhard}}, \bibnamefont{and}
  \bibinfo{author}{\bibfnamefont{J.}~\bibnamefont{Wrachtrup}},
  \bibinfo{journal}{Phys. Rev. Lett.} \textbf{\bibinfo{volume}{111}},
  \bibinfo{pages}{170801} (\bibinfo{year}{2013}).

\bibitem[{\citenamefont{Rybak et~al.}(2011)\citenamefont{Rybak, Amaran, Levin,
  Tomza, Moszynski, Kosloff, Koch, and Amitay}}]{RybakPRL11}
\bibinfo{author}{\bibfnamefont{L.}~\bibnamefont{Rybak}},
  \bibinfo{author}{\bibfnamefont{S.}~\bibnamefont{Amaran}},
  \bibinfo{author}{\bibfnamefont{L.}~\bibnamefont{Levin}},
  \bibinfo{author}{\bibfnamefont{M.}~\bibnamefont{Tomza}},
  \bibinfo{author}{\bibfnamefont{R.}~\bibnamefont{Moszynski}},
  \bibinfo{author}{\bibfnamefont{R.}~\bibnamefont{Kosloff}},
  \bibinfo{author}{\bibfnamefont{C.~P.} \bibnamefont{Koch}}, \bibnamefont{and}
  \bibinfo{author}{\bibfnamefont{Z.}~\bibnamefont{Amitay}},
  \bibinfo{journal}{Phys. Rev. Lett.} \textbf{\bibinfo{volume}{107}},
  \bibinfo{pages}{273001} (\bibinfo{year}{2011}).

\bibitem[{\citenamefont{Gonz\'alez-F\'erez and Koch}(2012)}]{GonzalezPRA12}
\bibinfo{author}{\bibfnamefont{R.}~\bibnamefont{Gonz\'alez-F\'erez}}
  \bibnamefont{and} \bibinfo{author}{\bibfnamefont{C.~P.} \bibnamefont{Koch}},
  \bibinfo{journal}{Phys. Rev. A} \textbf{\bibinfo{volume}{86}},
  \bibinfo{pages}{063420} (\bibinfo{year}{2012}).

\bibitem[{\citenamefont{Rice and Zhao}(2000)}]{RiceBook}
\bibinfo{author}{\bibfnamefont{S.~A.} \bibnamefont{Rice}} \bibnamefont{and}
  \bibinfo{author}{\bibfnamefont{M.}~\bibnamefont{Zhao}},
  \emph{\bibinfo{title}{Optical control of molecular dynamics}}
  (\bibinfo{publisher}{John Wiley \& Sons}, \bibinfo{year}{2000}).

\bibitem[{\citenamefont{Brumer and Shapiro}(2003)}]{ShapiroBook}
\bibinfo{author}{\bibfnamefont{P.}~\bibnamefont{Brumer}} \bibnamefont{and}
  \bibinfo{author}{\bibfnamefont{M.}~\bibnamefont{Shapiro}},
  \emph{\bibinfo{title}{Principles and Applications of the Quantum Control of
  Molecular Processes}} (\bibinfo{publisher}{Wiley Interscience},
  \bibinfo{year}{2003}).

\bibitem[{\citenamefont{Tannor et~al.}(1992)\citenamefont{Tannor, Kazakov, and
  Orlov}}]{Tannor92}
\bibinfo{author}{\bibfnamefont{D.}~\bibnamefont{Tannor}},
  \bibinfo{author}{\bibfnamefont{V.}~\bibnamefont{Kazakov}}, \bibnamefont{and}
  \bibinfo{author}{\bibfnamefont{V.}~\bibnamefont{Orlov}}, in
  \emph{\bibinfo{booktitle}{Time-dependent quantum molecular dynamics}}, edited
  by \bibinfo{editor}{\bibfnamefont{J.}~\bibnamefont{Broeckhove}}
  \bibnamefont{and}
  \bibinfo{editor}{\bibfnamefont{L.}~\bibnamefont{Lathouwers}}
  (\bibinfo{publisher}{Plenum}, \bibinfo{year}{1992}), pp.
  \bibinfo{pages}{347--360}.

\bibitem[{\citenamefont{Werschnik and Gross}(2007)}]{WerschnikJPB07}
\bibinfo{author}{\bibfnamefont{J.}~\bibnamefont{Werschnik}} \bibnamefont{and}
  \bibinfo{author}{\bibfnamefont{E.~K.~U.} \bibnamefont{Gross}},
  \bibinfo{journal}{J. Phys. B} \textbf{\bibinfo{volume}{40}},
  \bibinfo{pages}{R175} (\bibinfo{year}{2007}).

\bibitem[{\citenamefont{Palao and Kosloff}(2002)}]{JosePRL02}
\bibinfo{author}{\bibfnamefont{J.~P.} \bibnamefont{Palao}} \bibnamefont{and}
  \bibinfo{author}{\bibfnamefont{R.}~\bibnamefont{Kosloff}},
  \bibinfo{journal}{Phys. Rev. Lett.} \textbf{\bibinfo{volume}{89}},
  \bibinfo{pages}{188301} (\bibinfo{year}{2002}).

\bibitem[{\citenamefont{Doria et~al.}(2011)\citenamefont{Doria, Calarco, and
  Montangero}}]{DoriaPRL11}
\bibinfo{author}{\bibfnamefont{P.}~\bibnamefont{Doria}},
  \bibinfo{author}{\bibfnamefont{T.}~\bibnamefont{Calarco}}, \bibnamefont{and}
  \bibinfo{author}{\bibfnamefont{S.}~\bibnamefont{Montangero}},
  \bibinfo{journal}{Phys. Rev. Lett.} \textbf{\bibinfo{volume}{106}},
  \bibinfo{pages}{190501} (\bibinfo{year}{2011}).

\bibitem[{\citenamefont{Kaiser and May}(2004)}]{KaiserJCP04}
\bibinfo{author}{\bibfnamefont{A.}~\bibnamefont{Kaiser}} \bibnamefont{and}
  \bibinfo{author}{\bibfnamefont{V.}~\bibnamefont{May}}, \bibinfo{journal}{J.
  Comp. Phys.} \textbf{\bibinfo{volume}{121}}, \bibinfo{pages}{2528}
  (\bibinfo{year}{2004}).

\bibitem[{\citenamefont{Caneva et~al.}(2011)\citenamefont{Caneva, Calarco, and
  Montangero}}]{CanevaPRA11}
\bibinfo{author}{\bibfnamefont{T.}~\bibnamefont{Caneva}},
  \bibinfo{author}{\bibfnamefont{T.}~\bibnamefont{Calarco}}, \bibnamefont{and}
  \bibinfo{author}{\bibfnamefont{S.}~\bibnamefont{Montangero}},
  \bibinfo{journal}{Phys. Rev. A} \textbf{\bibinfo{volume}{84}},
  \bibinfo{pages}{022326} (\bibinfo{year}{2011}).

\bibitem[{\citenamefont{Konnov and Krotov}(1999)}]{Konnov99}
\bibinfo{author}{\bibfnamefont{A.}~\bibnamefont{Konnov}} \bibnamefont{and}
  \bibinfo{author}{\bibfnamefont{V.}~\bibnamefont{Krotov}},
  \bibinfo{journal}{Automation and Remote Control}
  \textbf{\bibinfo{volume}{60}}, \bibinfo{pages}{1427} (\bibinfo{year}{1999}).

\bibitem[{\citenamefont{Sklarz and Tannor}(2002)}]{sklarz:02}
\bibinfo{author}{\bibfnamefont{S.~E.} \bibnamefont{Sklarz}} \bibnamefont{and}
  \bibinfo{author}{\bibfnamefont{D.~J.} \bibnamefont{Tannor}},
  \bibinfo{journal}{Phys. Rev. A} \textbf{\bibinfo{volume}{66}},
  \bibinfo{pages}{053619} (\bibinfo{year}{2002}).

\bibitem[{\citenamefont{Khaneja et~al.}(2005)\citenamefont{Khaneja, Reiss,
  Kehlet, Schulte-Herbr\"uggen, and Glaser}}]{KhanejaJMR05}
\bibinfo{author}{\bibfnamefont{N.}~\bibnamefont{Khaneja}},
  \bibinfo{author}{\bibfnamefont{T.}~\bibnamefont{Reiss}},
  \bibinfo{author}{\bibfnamefont{C.}~\bibnamefont{Kehlet}},
  \bibinfo{author}{\bibfnamefont{T.}~\bibnamefont{Schulte-Herbr\"uggen}},
  \bibnamefont{and} \bibinfo{author}{\bibfnamefont{S.~J.}
  \bibnamefont{Glaser}}, \bibinfo{journal}{J. Mar. Res.}
  \textbf{\bibinfo{volume}{172}}, \bibinfo{pages}{296 } (\bibinfo{year}{2005}).

\bibitem[{\citenamefont{Machnes et~al.}(2011)\citenamefont{Machnes, Sander,
  Glaser, de~Fouqui\`eres, Gruslys, Schirmer, and
  Schulte-Herbr\"uggen}}]{MachnesPRA11}
\bibinfo{author}{\bibfnamefont{S.}~\bibnamefont{Machnes}},
  \bibinfo{author}{\bibfnamefont{U.}~\bibnamefont{Sander}},
  \bibinfo{author}{\bibfnamefont{S.~J.} \bibnamefont{Glaser}},
  \bibinfo{author}{\bibfnamefont{P.}~\bibnamefont{de~Fouqui\`eres}},
  \bibinfo{author}{\bibfnamefont{A.}~\bibnamefont{Gruslys}},
  \bibinfo{author}{\bibfnamefont{S.}~\bibnamefont{Schirmer}}, \bibnamefont{and}
  \bibinfo{author}{\bibfnamefont{T.}~\bibnamefont{Schulte-Herbr\"uggen}},
  \bibinfo{journal}{Phys. Rev. A} \textbf{\bibinfo{volume}{84}},
  \bibinfo{pages}{022305} (\bibinfo{year}{2011}).

\bibitem[{\citenamefont{Eitan et~al.}(2012)\citenamefont{Eitan, Mundt, and
  Tannor}}]{eitan:12}
\bibinfo{author}{\bibfnamefont{R.}~\bibnamefont{Eitan}},
  \bibinfo{author}{\bibfnamefont{M.}~\bibnamefont{Mundt}}, \bibnamefont{and}
  \bibinfo{author}{\bibfnamefont{D.~J.} \bibnamefont{Tannor}},
  \bibinfo{journal}{Phys. Rev. A} \textbf{\bibinfo{volume}{83}},
  \bibinfo{pages}{053426} (\bibinfo{year}{2012}).

\bibitem[{\citenamefont{Caneva et~al.}(2009)\citenamefont{Caneva, Murphy,
  Calarco, Fazio, Montangero, Giovannetti, and Santoro}}]{CanevaPRL09}
\bibinfo{author}{\bibfnamefont{T.}~\bibnamefont{Caneva}},
  \bibinfo{author}{\bibfnamefont{M.}~\bibnamefont{Murphy}},
  \bibinfo{author}{\bibfnamefont{T.}~\bibnamefont{Calarco}},
  \bibinfo{author}{\bibfnamefont{R.}~\bibnamefont{Fazio}},
  \bibinfo{author}{\bibfnamefont{S.}~\bibnamefont{Montangero}},
  \bibinfo{author}{\bibfnamefont{V.}~\bibnamefont{Giovannetti}},
  \bibnamefont{and} \bibinfo{author}{\bibfnamefont{G.~E.}
  \bibnamefont{Santoro}}, \bibinfo{journal}{Phys. Rev. Lett.}
  \textbf{\bibinfo{volume}{103}}, \bibinfo{pages}{240501}
  (\bibinfo{year}{2009}).

\bibitem[{\citenamefont{Hohenester et~al.}(2007)\citenamefont{Hohenester,
  Rekdal, Borzi, and Schmiedmayer}}]{hohenester.pra:07}
\bibinfo{author}{\bibfnamefont{U.}~\bibnamefont{Hohenester}},
  \bibinfo{author}{\bibfnamefont{P.~K.} \bibnamefont{Rekdal}},
  \bibinfo{author}{\bibfnamefont{A.}~\bibnamefont{Borzi}}, \bibnamefont{and}
  \bibinfo{author}{\bibfnamefont{J.}~\bibnamefont{Schmiedmayer}},
  \bibinfo{journal}{Phys. Rev. A} \textbf{\bibinfo{volume}{75}},
  \bibinfo{pages}{023602} (\bibinfo{year}{2007}).

\bibitem[{\citenamefont{Hohenester}(2014)}]{hohenester.cpc:14a}
\bibinfo{author}{\bibfnamefont{U.}~\bibnamefont{Hohenester}},
  \bibinfo{journal}{Comp. Phys. Commun.} \textbf{\bibinfo{volume}{185}},
  \bibinfo{pages}{194} (\bibinfo{year}{2014}).

\bibitem[{\citenamefont{Reich et~al.}(2012)\citenamefont{Reich, Ndong, and
  Koch}}]{reich:12}
\bibinfo{author}{\bibfnamefont{D.~M.} \bibnamefont{Reich}},
  \bibinfo{author}{\bibfnamefont{M.}~\bibnamefont{Ndong}}, \bibnamefont{and}
  \bibinfo{author}{\bibfnamefont{C.~P.} \bibnamefont{Koch}},
  \bibinfo{journal}{J. Chem Phys.} \textbf{\bibinfo{volume}{136}},
  \bibinfo{pages}{104103} (\bibinfo{year}{2012}).

\bibitem[{\citenamefont{Shin et~al.}(2004)\citenamefont{Shin, Saba, Pasquini,
  Ketterle, Pritchard, and Leanhardt}}]{shin:04}
\bibinfo{author}{\bibfnamefont{Y.}~\bibnamefont{Shin}},
  \bibinfo{author}{\bibfnamefont{M.}~\bibnamefont{Saba}},
  \bibinfo{author}{\bibfnamefont{T.~A.} \bibnamefont{Pasquini}},
  \bibinfo{author}{\bibfnamefont{W.}~\bibnamefont{Ketterle}},
  \bibinfo{author}{\bibfnamefont{D.~E.} \bibnamefont{Pritchard}},
  \bibnamefont{and} \bibinfo{author}{\bibfnamefont{A.~E.}
  \bibnamefont{Leanhardt}}, \bibinfo{journal}{Phys. Rev. Lett.}
  \textbf{\bibinfo{volume}{92}}, \bibinfo{pages}{050405}
  (\bibinfo{year}{2004}).

\bibitem[{\citenamefont{Schumm et~al.}(2005)\citenamefont{Schumm, Hofferberth,
  Andersson, Wildermuth, Groth, Bar-Joseph, Schmiedmayer, and
  Kr\"uger}}]{schumm:05}
\bibinfo{author}{\bibfnamefont{T.}~\bibnamefont{Schumm}},
  \bibinfo{author}{\bibfnamefont{S.}~\bibnamefont{Hofferberth}},
  \bibinfo{author}{\bibfnamefont{L.~M.} \bibnamefont{Andersson}},
  \bibinfo{author}{\bibfnamefont{S.}~\bibnamefont{Wildermuth}},
  \bibinfo{author}{\bibfnamefont{S.}~\bibnamefont{Groth}},
  \bibinfo{author}{\bibfnamefont{I.}~\bibnamefont{Bar-Joseph}},
  \bibinfo{author}{\bibfnamefont{J.}~\bibnamefont{Schmiedmayer}},
  \bibnamefont{and} \bibinfo{author}{\bibfnamefont{P.}~\bibnamefont{Kr\"uger}},
  \bibinfo{journal}{Nature Phys.} \textbf{\bibinfo{volume}{1}},
  \bibinfo{pages}{57} (\bibinfo{year}{2005}).

\bibitem[{\citenamefont{Grond et~al.}(2010)\citenamefont{Grond, Hohenester,
  Mazets, and Schmiedmayer}}]{grond.njp:10}
\bibinfo{author}{\bibfnamefont{J.}~\bibnamefont{Grond}},
  \bibinfo{author}{\bibfnamefont{U.}~\bibnamefont{Hohenester}},
  \bibinfo{author}{\bibfnamefont{I.}~\bibnamefont{Mazets}}, \bibnamefont{and}
  \bibinfo{author}{\bibfnamefont{J.}~\bibnamefont{Schmiedmayer}},
  \bibinfo{journal}{New J. Phys.} \textbf{\bibinfo{volume}{12}},
  \bibinfo{pages}{065036} (\bibinfo{year}{2010}).

\bibitem[{\citenamefont{Peirce et~al.}(1988)\citenamefont{Peirce, Dahleh, and
  Rabitz}}]{peirce:88}
\bibinfo{author}{\bibfnamefont{A.~P.} \bibnamefont{Peirce}},
  \bibinfo{author}{\bibfnamefont{M.~A.} \bibnamefont{Dahleh}},
  \bibnamefont{and} \bibinfo{author}{\bibfnamefont{H.}~\bibnamefont{Rabitz}},
  \bibinfo{journal}{Phys. Rev. A} \textbf{\bibinfo{volume}{37}},
  \bibinfo{pages}{4950} (\bibinfo{year}{1988}).

\bibitem[{\citenamefont{B{\"u}cker et~al.}(2013)\citenamefont{B{\"u}cker,
  Berrada, va{n Frank}, Schumm, Schaff, Schmiedmayer, J{\"a}ger, Grond, and
  Hohenester}}]{buecker:13}
\bibinfo{author}{\bibfnamefont{R.}~\bibnamefont{B{\"u}cker}},
  \bibinfo{author}{\bibfnamefont{T.}~\bibnamefont{Berrada}},
  \bibinfo{author}{\bibfnamefont{S.}~\bibnamefont{va{n Frank}}},
  \bibinfo{author}{\bibfnamefont{T.}~\bibnamefont{Schumm}},
  \bibinfo{author}{\bibfnamefont{J.~F.} \bibnamefont{Schaff}},
  \bibinfo{author}{\bibfnamefont{J.}~\bibnamefont{Schmiedmayer}},
  \bibinfo{author}{\bibfnamefont{G.}~\bibnamefont{J{\"a}ger}},
  \bibinfo{author}{\bibfnamefont{J.}~\bibnamefont{Grond}}, \bibnamefont{and}
  \bibinfo{author}{\bibfnamefont{U.}~\bibnamefont{Hohenester}},
  \bibinfo{journal}{J. Phys. B} \textbf{\bibinfo{volume}{46}},
  \bibinfo{pages}{104012} (\bibinfo{year}{2013}).

\bibitem[{\citenamefont{Leggett}(2001)}]{leggett:01}
\bibinfo{author}{\bibfnamefont{A.}~\bibnamefont{Leggett}},
  \bibinfo{journal}{Rev. Mod. Phys.} \textbf{\bibinfo{volume}{73}},
  \bibinfo{pages}{307} (\bibinfo{year}{2001}).

\bibitem[{\citenamefont{Grond et~al.}(2009)\citenamefont{Grond, von Winckel,
  Schmiedmayer, and Hohenester}}]{grond.pra:09b}
\bibinfo{author}{\bibfnamefont{J.}~\bibnamefont{Grond}},
  \bibinfo{author}{\bibfnamefont{G.}~\bibnamefont{von Winckel}},
  \bibinfo{author}{\bibfnamefont{J.}~\bibnamefont{Schmiedmayer}},
  \bibnamefont{and}
  \bibinfo{author}{\bibfnamefont{U.}~\bibnamefont{Hohenester}},
  \bibinfo{journal}{Phys. Rev. A} \textbf{\bibinfo{volume}{80}},
  \bibinfo{pages}{053625} (\bibinfo{year}{2009}).

\bibitem[{\citenamefont{J\"ager and Hohenester}(2013)}]{jaeger.pra:13}
\bibinfo{author}{\bibfnamefont{G.}~\bibnamefont{J\"ager}} \bibnamefont{and}
  \bibinfo{author}{\bibfnamefont{U.}~\bibnamefont{Hohenester}},
  \bibinfo{journal}{Phys. Rev. A} \textbf{\bibinfo{volume}{88}},
  \bibinfo{pages}{035601} (\bibinfo{year}{2013}).

\bibitem[{\citenamefont{Borzi and Hohenester}(2008)}]{borzi:08}
\bibinfo{author}{\bibfnamefont{A.}~\bibnamefont{Borzi}} \bibnamefont{and}
  \bibinfo{author}{\bibfnamefont{U.}~\bibnamefont{Hohenester}},
  \bibinfo{journal}{SIAM Journal of Scientific Computing}
  \textbf{\bibinfo{volume}{30}}, \bibinfo{pages}{441} (\bibinfo{year}{2008}).

\bibitem[{\citenamefont{v{on Winckel} and Borzi}(2008)}]{vonwinckel:08}
\bibinfo{author}{\bibfnamefont{G.}~\bibnamefont{v{on Winckel}}}
  \bibnamefont{and} \bibinfo{author}{\bibfnamefont{A.}~\bibnamefont{Borzi}},
  \bibinfo{journal}{Inverse Problems} \textbf{\bibinfo{volume}{24}},
  \bibinfo{pages}{034007} (\bibinfo{year}{2008}).

\bibitem[{\citenamefont{Dion and Cances}(2007)}]{dion:07}
\bibinfo{author}{\bibfnamefont{C.~M.} \bibnamefont{Dion}} \bibnamefont{and}
  \bibinfo{author}{\bibfnamefont{E.}~\bibnamefont{Cances}},
  \bibinfo{journal}{Comp. Phys. Commun.} \textbf{\bibinfo{volume}{177}},
  \bibinfo{pages}{787} (\bibinfo{year}{2007}).

\bibitem[{\citenamefont{Bertsekas}(1999)}]{bertsekas:99}
\bibinfo{author}{\bibfnamefont{D.~P.} \bibnamefont{Bertsekas}},
  \emph{\bibinfo{title}{Nonlinear Programming}} (\bibinfo{publisher}{Athena
  Scientific}, \bibinfo{address}{Cambridge, UK}, \bibinfo{year}{1999}).

\bibitem[{\citenamefont{Palao and Kosloff}(2003)}]{PalaoPRA03}
\bibinfo{author}{\bibfnamefont{J.~P.} \bibnamefont{Palao}} \bibnamefont{and}
  \bibinfo{author}{\bibfnamefont{R.}~\bibnamefont{Kosloff}},
  \bibinfo{journal}{Phys. Rev. A} \textbf{\bibinfo{volume}{68}},
  \bibinfo{pages}{062308} (\bibinfo{year}{2003}).

\bibitem[{\citenamefont{Lesanovsky et~al.}(2006)\citenamefont{Lesanovsky,
  Schumm, Hofferberth, Andersson, Kr\"uger, and Schmiedmayer}}]{lesanovsky:06}
\bibinfo{author}{\bibfnamefont{I.}~\bibnamefont{Lesanovsky}},
  \bibinfo{author}{\bibfnamefont{T.}~\bibnamefont{Schumm}},
  \bibinfo{author}{\bibfnamefont{S.}~\bibnamefont{Hofferberth}},
  \bibinfo{author}{\bibfnamefont{L.~M.} \bibnamefont{Andersson}},
  \bibinfo{author}{\bibfnamefont{P.}~\bibnamefont{Kr\"uger}}, \bibnamefont{and}
  \bibinfo{author}{\bibfnamefont{J.}~\bibnamefont{Schmiedmayer}},
  \bibinfo{journal}{Phys. Rev. A} \textbf{\bibinfo{volume}{73}},
  \bibinfo{eid}{033619} (\bibinfo{year}{2006}).

\bibitem[{\citenamefont{Palao et~al.}(2013)\citenamefont{Palao, Reich, and
  Koch}}]{JosePRA13}
\bibinfo{author}{\bibfnamefont{J.~P.} \bibnamefont{Palao}},
  \bibinfo{author}{\bibfnamefont{D.~M.} \bibnamefont{Reich}}, \bibnamefont{and}
  \bibinfo{author}{\bibfnamefont{C.~P.} \bibnamefont{Koch}},
  \bibinfo{journal}{Phys. Rev. A} \textbf{\bibinfo{volume}{88}},
  \bibinfo{pages}{053409} (\bibinfo{year}{2013}).

\end{thebibliography}

\end{document}